\documentclass{article}
\usepackage[utf8]{inputenc}
\usepackage{authblk}
\usepackage{setspace}
\usepackage[margin=1.25in]{geometry}
\usepackage{graphicx}
\graphicspath{ {./figures/} }
\usepackage{subcaption}
\usepackage{booktabs}
\usepackage{amsmath}
\usepackage{lineno}
\usepackage{graphicx}
\usepackage{dcolumn}
\usepackage{bm}
\usepackage{array}
\usepackage{footnote}
\usepackage{amssymb}
\usepackage{listings}
\usepackage{xcolor}
\usepackage{hyperref}

\usepackage[style=nejm, 
citestyle=numeric-comp,
sorting=none]{biblatex}
\title{Towards Efficient and Accurate EMRI Parameter Estimation: A Machine Learning-Enhanced MCMC Framework}

\author[1$\dag$,2]{Bo Liang}
\author[1$\dag$,2,3]{Chang Liu}
\author[4]{Hanlin Song}
\author[5]{Zhenwei Lyu}
\author[1*]{Minghui Du}
\author[1*,2,6]{Peng Xu}
\author[1,2,7]{Ziren Luo}
\author[8]{Sensen He}
\author[8]{Haohao Gu}
\author[1]{Tianyu Zhao}
\author[1]{Manjia Liang}
\author[1]{Yuxiang Xu}
\author[3]{Li-e Qiang}
\author[9]{Mingming Sun}
\author[10]{Wei-Liang Qian}

\affil[1]{Center for Gravitational Wave Experiment, National Microgravity Laboratory, Institute of Mechanics, Chinese Academy of Sciences, Beijing 100190, China}
\affil[2]{Taiji Laboratory for Gravitational Wave Universe (Beijing/Hangzhou), University of Chinese Academy of Sciences (UCAS), Beijing 100049, China}
\affil[3]{National Space Science Center, Chinese Academy of Sciences, Beijing 100190, China}
\affil[4]{School of Physics, Peking University, Beijing 100871, China}
\affil[5]{Leicester International Institute, Dalian University of Technology, Panjin 124221, China}
\affil[6]{Lanzhou Center of Theoretical Physics, Lanzhou University, Lanzhou 730000, China}
\affil[7]{Key Laboratory of Gravitational Wave Precision Measurement of Zhejiang Province, Hangzhou Institute for Advanced Study, UCAS, Hangzhou 310024, China}
\affil[8]{Baidu Inc., Beijing 100085, P. R. China}
\affil[9]{AGI Lab, Beijing Institute of Mathematical Sciences and Applications, Beijing, China}
\affil[10]{Escola de Engenharia de Lorena, Universidade de São Paulo, 12602-810, Lorena, SP, Brazil}
\affil[*]{duminghui@imech.ac.cn, xupeng@imech.ac.cn}
\affil[$\dag$]{These authors contributed equally to this work.}

\date{}

\begin{document}

\maketitle

\begin{abstract}
The detection of gravitational waves from extreme-mass-ratio inspirals (EMRIs) in space-based antennas like Taiji and LISA promises deep insights into strong-field gravity and black hole physics. However, the complex, highly degenerate, and non-convex likelihood landscapes characteristic of EMRI parameter spaces pose severe challenges for conventional Markov chain Monte Carlo (MCMC) methods. 
Under realistic instrumental noise and broad priors, these methods demand impractical computational costs but are prone to becoming trapped in local maxima, leading to biased and unreliable parameter estimates. 
To address these challenges, we introduce Flow-Matching Markov Chain Monte Carlo (FM-MCMC), a novel Bayesian framework that integrates continuous normalizing flows (CNFs) with parallel tempering MCMC (PTMCMC). 
By generating high-likelihood regions via CNFs and refining them through PTMCMC, FM-MCMC enables robust exploration of the nontrivial parameter spaces, achieves orders-of-magnitude improvement in computational efficiency,  and more importantly, ensures statistically unbiased inference.
By enabling real-time, unbiased parameter inference, FM-MCMC could unlock the full scientific potential of EMRI observations, and would serve as a scalable pipeline for precision gravitational-wave astronomy.

\end{abstract}


\section{Introduction}\label{Sec:intro}

Gravitational wave (GW) observation has gradually revolutionized our view of the Universe since the first detection of GW150914 by the LIGO-Virgo Collaboration~\cite{LIGOScientific:2014pky, LIGOScientific:2016aoc}. 
Over the past decade, ground-based detectors such as LIGO, Virgo, and KAGRA have cataloged over one hundred merger events of compact binaries, including binary black holes, binary neutron stars, and black hole-neutron star systems~\cite{LIGO2025, LIGOScientific:2016aoc, LIGOScientific:2020ibl}. 
These observatories, limited mainly by terrestrial noises, operate in the $10 \sim 10^3$ Hz frequency band and probe astrophysical systems with masses up to hundreds of solar masses~\cite{KAGRA:2021vkt,doi:10.34133/2021/6014164,doi:10.34133/research.0302}.
However, the millihertz GW universe encompassing sources like extreme-mass-ratio inspirals (EMRIs), massive black hole binaries, and galactic binaries remains largely unexplored. 
The planned space-based interferometers, such as Taiji, TianQin and LISA~\cite{ Hu:2017mde, TianQin:2015yph, LISA:2017pwj}, promise to open a new window into such low-frequency GW universe, in which EMRIs are among the most promising sources for probing strong-field gravity and black hole physics \cite{Babak:2017tow, Gair:2012nm}.

EMRIs, in which a stellar-mass compact object inspirals into a supermassive black hole through more than $10^5$ orbital cycles, encode a wealth of information about the spacetime geometry and accretion dynamics in the vicinity of the supermassive black hole \cite{Laghi:2021pqk}. 
These systems are expected to retain significant orbital eccentricities in the final stages before plunge \cite{Babak_2017}. 
The combination of high eccentricity and extreme mass ratio produces GW signals with a rich harmonic structure and tens of thousands of orbital cycles within the detector’s sensitive frequency band.
Such unique waveform characteristics make EMRIs powerful probes of the nature of gravity (including, for example, the direct test of the ``no-hair'' theorem)~\cite{Sopuerta_2009, Barack_2007, speri2024probingfundamentalphysicsextreme,zou2025constraintdeviationkerrmetric, PhysRevD.100.084055} and of the astrophysical environments (including dark matter distributions) surrounding the supermassive black holes ~\cite{Gair_2010, Speri_2023, cole2022disksspikescloudsdistinguishing, PhysRevD.102.103022, Yue_2019}. 
While, unlocking EMRI's full scientific objectives and implications will require precise measurements of key physical parameters, such as the central black hole's mass $M$ and spin $a$, the mass of the stellar-mass compact object  $\mu$ and the orbital eccentricity $e$.

However, the application of traditional Bayesian methods like Markov Chain Monte Carlo (MCMC) to EMRI signal parameter estimation faces significant challenges, primarily due to prohibitive computational costs and inadequate exploration of the parameter space~\cite{Gair:2004iv, saltas2023emri_mc}.
MCMC relies on intensive waveform evaluations to explore the parameter space,
as the Bayesian posterior occupies only a minuscule region within the high-dimensional space~\cite{Chua:2020stf}, necessitating an enormous number of iterations to locate the global optimum corresponding to the true parameters. 
Such challenges are further exacerbated by the long duration of EMRI signals (lasting months to years~\cite{Gair:2004iv, saltas2023emri_mc}),  which significantly increases the computational cost of each likelihood evaluation.
Another fundamental challenge in EMRI searches stems from the highly non-convex structure of the likelihood surface, which is riddled with numerous local maxima~\cite{MockLISADataChallengeTaskForce:2009wir,Chua:2021aah,universe10020096}.
These local peaks correspond to non-local degeneracies in the parameter space, that distinct regions could yield waveforms nearly indistinguishable from the true signal.  
In other words, widely separated parameter regions can produce highly similar gravitational waveforms, leading to multiple strong but spurious likelihood peaks. Crucially, the global maximum associated with the true parameters often lies far from these suboptimal solutions in parameter space, making it extremely difficult to identify without exhaustive exploration. 
Traditional methods like grid searches or MCMC are highly susceptible to becoming trapped in these local maxima during the parameter estimation  process~\cite{Chua:2020stf}.
\begin{figure*}[ht!]
  \centering
  \includegraphics[width=1.0\textwidth]{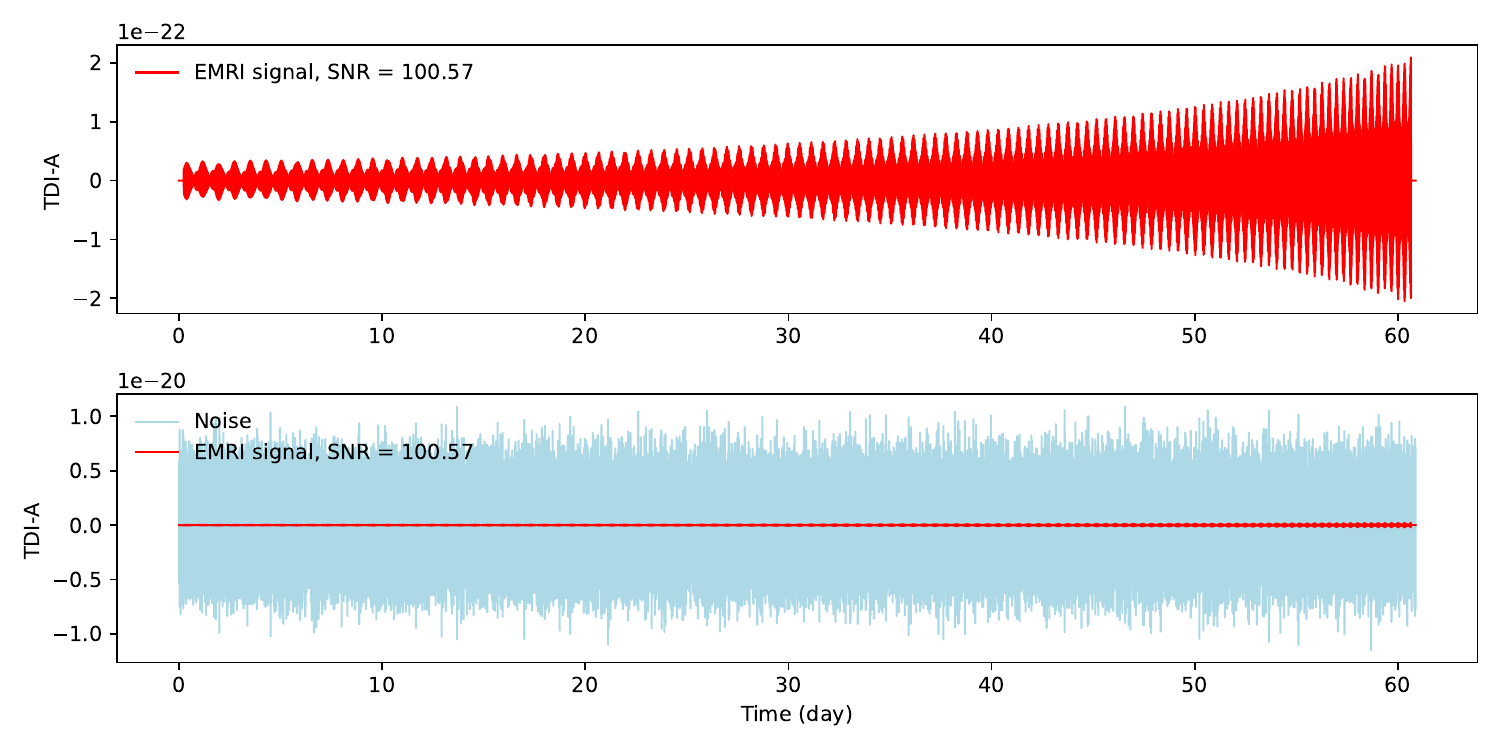}
  \caption{ (Top) The EMRI signal  under instrumental noise conditions. The  EMRI waveform (red curve) is generated with the augmented analytic kludge model, showing the Time-Delay Interferometry channel A observable over a 60 days observation period. 
(Bottom) The same EMRI signal combined with simulated Taiji instrumental noise (blue shading), illustrating the challenge of parameter recovery in realistic noise environments. 
  }
  \label{fig:TD}
\end{figure*}

\begin{figure*}[ht!]
  \centering
  \includegraphics[width=1.0\textwidth]{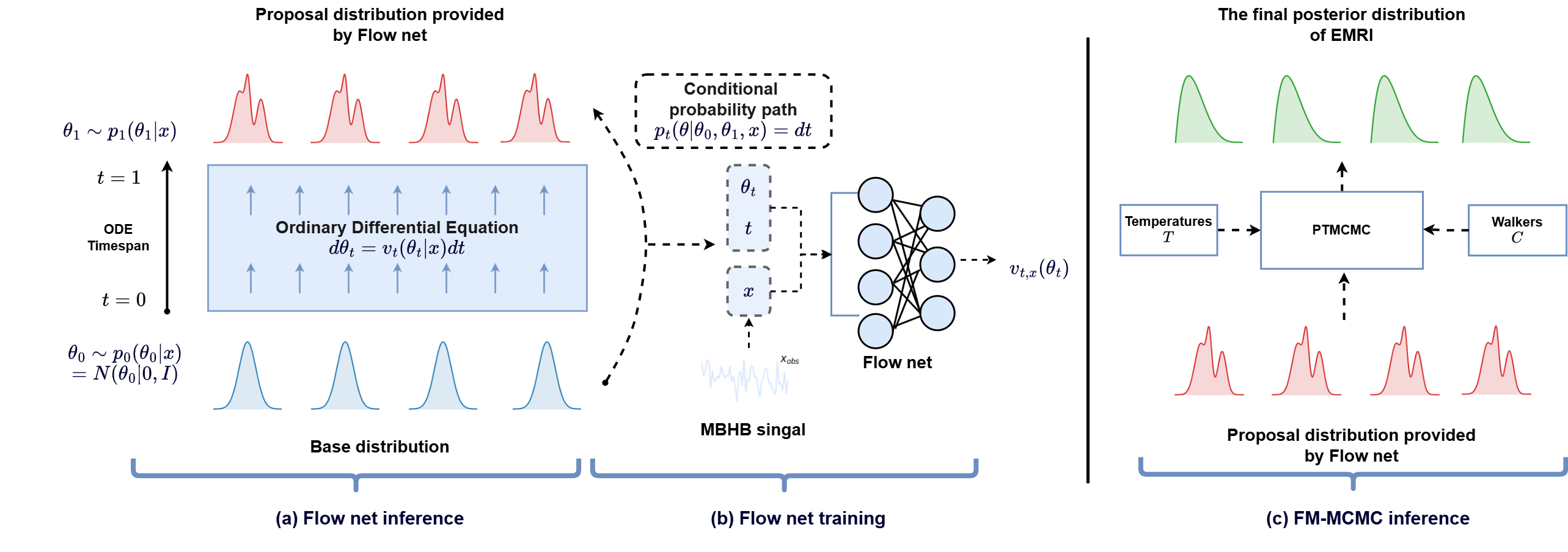}
  \caption{
(a) Flow net inference stage. Demonstrates the transformation from a baseline Gaussian distribution \( \theta_0 \sim p_{\theta_0}(\theta_0|x) = \mathcal{N}(\theta_0|0,I) \) (blue) to the time-evolving EMRI posterior distribution \( \theta_t \sim p_{\theta_t}(\theta_t|x) \) (pink) through an ordinary differential equation (ODE). 
Here, $\theta_0 \sim \mathcal{N}(0,I)$ denotes the baseline distribution, $x$ represents preprocessed EMRI data (details explained in Section~\ref{Sec:model}.), and $\theta_t$ models the continuously interpolated transformation state at normalized time $t\in[0,1]$.  
The conditional probability path of the entire ODE is modeled by Flow net.
The theoretical foundation and implementation details are comprehensively described in Subsection~\ref{cnf}.
(b) Flow net training stage. Illustrates the Flow net training process, where the Flow net learns the velocity field $v_{\theta}$ by minimizing the flow matching~\cite{Dax2023FlowMF} loss function.
(c) FM-MCMC inference stage. 
During FM-MCMC inference initialization, PTMCMC determines starting distribution points through its temperature parameter T and walks C. This initialization distribution is then generated by Flow net's proposal distribution. Final posterior sampling is executed exclusively via PTMCMC's likelihood evaluations, completing the Bayesian inference cycle for EMRI parameter estimation.
}
  \label{fig:model}
\end{figure*}

{
In data analysis scenarios, where instrumental noise is present (as illustrated in the bottom panel of Figure 1), the performance of traditional inference methods is further compromised~\cite{Cornish_2011, Babak_2009, ZOU_181, Chua:2021aah}. 
Instrumental noise introduces additional complexity to the likelihood surface, substantially increasing the number of local maxima. 
These maxima are not only more numerous but also sharper and more densely packed, making the parameter space highly rugged~\cite{ZOU_181}. 
As a result, inference algorithms such as MCMC or hybrid methods are prone to becoming trapped in these spurious local maxima, particularly when degeneracies in EMRI signals are present~\cite{Chua:2021aah}. 
This dramatically increases the difficulty of converging to  the true parameters. }

Key scientific objectives, such as testing strong-field gravity with EMRI systems or inferring black hole physical parameters, remain beyond reach with existing analysis techniques.  
Biased parameter estimation introduces significant systematic errors and may lead to incorrect astrophysical interpretations, e.g. misestimation of spin or orbital dynamics.  
Therefore, the development of novel algorithms capable of efficiently exploring high-dimensional, noise-corrupted parameter spaces while circumventing local optima is essential for fully unlocking the scientific opportunities offered by EMRI gravitational-wave observations.

To address these challenges, we propose Flow-Matching Markov Chain Monte Carlo (FM-MCMC): a hybrid framework  that integrates continuous normalizing flows (CNFs)~\cite{Dax2023FlowMF,lipman2022flow} with parallel tempering MCMC (PTMCMC)~\cite{Karnesis:2023ras, 2013PASP..125..306F}.
The FM-MCMC framework begins with flow matching, which rapidly and coarsely explores the parameter space through gradient-based trajectory learning. This approach effectively mitigates initialization sensitivity by identifying high-likelihood regions in a computationally efficient manner.
Subsequently, PTMCMC is initialized using samples from these high-likelihood regions, enabling fine-grained exploration and leveraging its robust capability for local precision posterior estimation.
Our framework is illustrated in Figure~\ref{fig:model}, with methodological details provided in Section~\ref{fmmc}.

For EMRI signals with signal-to-noise ratios (SNRs) exceeding 60—which defines typical ``bright'' sources prioritized in space-based detector observations—our method achieves reliable parameter recovery (especially for intrinsic parameters) with true values falling within the $1\sigma$ posterior credible intervals even under realistic instrumental noise conditions. In contrast, under identical noise conditions, traditional Bayesian methods such as standard MCMC sampling become trapped in local maxima of the likelihood function, resulting in substantial biases across all parameter estimates.
This methodological advance enables efficient and unbiased parameter estimation for EMRI systems under near-realistic observational conditions, addressing a long-standing bottleneck for conventional Bayesian inference approaches in EMRI analysis. 
With this breakthrough, our approach would pave the way for unlocking the scientific landscape revealed by EMRI observations.

The paper is structured as follows. 
Section~\ref{result} reports the results based on numerical simulations and compares the performance of our method with that of standard MCMC. 
In Section~\ref{sec:Conclusion}, we conclude with a discussion of implications and future research directions.
Section~\ref{Sec:model} details the methodologies of data generation and pre-processing, which support model training and validation. 
We then present the machine learning framework designed for Bayesian posterior estimation of EMRI signals. 

\section{Results}\label{result}
We present parameter estimation results for injected EMRI signals, with the source parameters drawn from the prior distributions detailed in Table~\ref{table:priors}. 
The waveform model employed in this study is based on EMRIs around a Kerr black hole, implemented within the \texttt{FEW} framework~\cite{Katz:2021yft,Chua:2020stf}. 
Detector response is modeled for the Taiji mission (as a representative of LISA-like missions), and the science data streams incorporate  Gaussian stationary noise generated according to Taiji’s noise budget. 

\begin{table*}[htb]
    \footnotesize
    \caption{Prior distributions used in this work. 
    For each parameter the table lists its lower and upper bounds, assuming a uniform distribution between them.
}
    \label{table:priors}
    \begin{tabular*}{\textwidth}{clccc}
    \toprule
        \hline
        \textbf{Parameter} & \textbf{Description} & \textbf{Prior Lower Bound} & \textbf{Prior Upper Bound}& \textbf{Units} \\
        \hline
        $M$ & Central  black hole mass &  $9 \times 10^5 $ & $1.1 \times 10^6 $ & $M_\odot$ \\
        $\mu$ &Secondary object mass & $50$  & $100$ & $M_\odot$  \\
        $a$ & Central black hole spin &  0.1 & 0.9 & --  \\
        $e_0$ &Orbital eccentricity &  0.1 & 0.6 & --  \\
        $\theta_S$ &Sky location polar angle &  0 rad & $\pi$ rad & rad \\
        $\phi_S$ &Sky location azimuthal angle &   0 rad & $2\pi$ rad & rad\\
        $\theta_K$ & Initial black hole spin polar angle &   0 rad & $\pi$ rad & rad \\
        $\phi_K$ & Initial black hole spin azimuthal angle & 0 rad  & $2\pi$ rad  & rad\\
        \hline
    \end{tabular*}
\end{table*}

{In this study, we focused on a subset of the parameters, especially the mass $M$ and spin $a$ of the central black hole, as well as 
the mass of the stellar-mass compact object  $\mu$ and the orbital eccentricity $e$, 
which are crucial for understanding the fundamental properties of the system and hold significant implications
for strong-field gravity and astrophysics~\cite{LISAWaveformWhitePaper}. 
We have simplified the analysis to an 8-dimensional parameter space to demonstrate the capabilities of our method. 
The full 14-dimensional parameter space, which includes orbital semi-latus rectum $p_0$ and luminosity distance $d$, will be addressed in future work.
As a representative example, the injected EMRI signal analyzed in this section has the following parameters: primary mass $M = 1.0593 \times 10^6 M_\odot$, secondary mass $\mu  = 8.6556 M_\odot$, dimensionless spin $a = 0.1931$, initial semi-latus rectum $p_{0} = 10.0$, initial eccentricity $e_0 = 0.4870$, luminosity distance = 2.0 Gpc, sky and spin orientations ($\theta_S$,  $\phi_S$, $\theta_K$, $\phi_K$)  = (1.672, 1.0372, 0.4040, 3.6169) radians, and the initial orbital phase is set to zero. The waveform is sampled at 25-second intervals over a two-month observation period.
The resulting signal-to-noise ratio (SNR) is 87.9.
}


To benchmark against our machine learning approach, we perform traditional Bayesian inference using \texttt{Eryn}~\cite{Karnesis_2023}, an advanced MCMC sampler built upon \texttt{emcee}~\cite{Foreman_Mackey_2013}.
\texttt{Eryn} has emerged as a popular tool for space-based GW data analysis, particularly in multi-source and high-dimensional inference problems (e.g., Ref.~\cite{PhysRevD.111.024060}).
We apply both \texttt{Eryn} and our method to the same injected EMRI signal, enabling a direct comparison of their sampling efficiencies, convergence properties, and parameter estimation accuracies.
For such comparison, each method employs two distinct initialization strategies:
\begin{itemize}
    \item [(1)] Informative initialization setting: Both intrinsic parameters ($M$, $\mu$, $a$, $e_0$) and extrinsic parameters ($\theta_S$,  $\phi_S$, $\theta_K$, $\phi_K$) are initialized within a narrow neighborhood of the true values, with absolute deviations less than $10^{-7}$ times the corresponding true parameter values. This configuration provides favorable initial conditions to assess convergence under idealized, truth-proximal starting points. 
    \item  [(2)] Practical initialization setting: The intrinsic parameters ($M$, $\mu$, $a$, $e_0$) are initialized by randomly drawing samples from their respective broad prior distributions, as defined in Table~\ref{table:priors}. In contrast, the extrinsic parameters ($\theta_S$, $\phi_S$, $\theta_K$, $\phi_K$) are initialized identically to setting (1), within a tight range ($< 10^{-7} \times$ true values) around the true values. This setup evaluates the sampler’s ability to recover the correct posterior mode when intrinsic parameters are initialized under realistic (i.e., uninformed) priors.
    
\end{itemize}
{Both settings employed identical parallel tempering configurations with 20 walkers and 20 temperature ladder levels. For FM-MCMC, 20,000 iterations were used, while PT-MCMC was run with 50,000 iterations in order to provide sufficient sampling. Despite this increased iteration count, PT-MCMC still struggled to achieve convergence, highlighting the efficiency advantage of FM-MCMC.
For FM-MCMC, the priors are fixed and generated by FMPE, which also provides initial sampling points. This setup reflects the different strategies of the two approaches: PT-MCMC depends on broad predefined priors, while FM-MCMC leverages informed priors from FMPE to improve efficiency.}

\begin{figure*}[htb]
  \centering
  \includegraphics[width=0.78\textwidth]{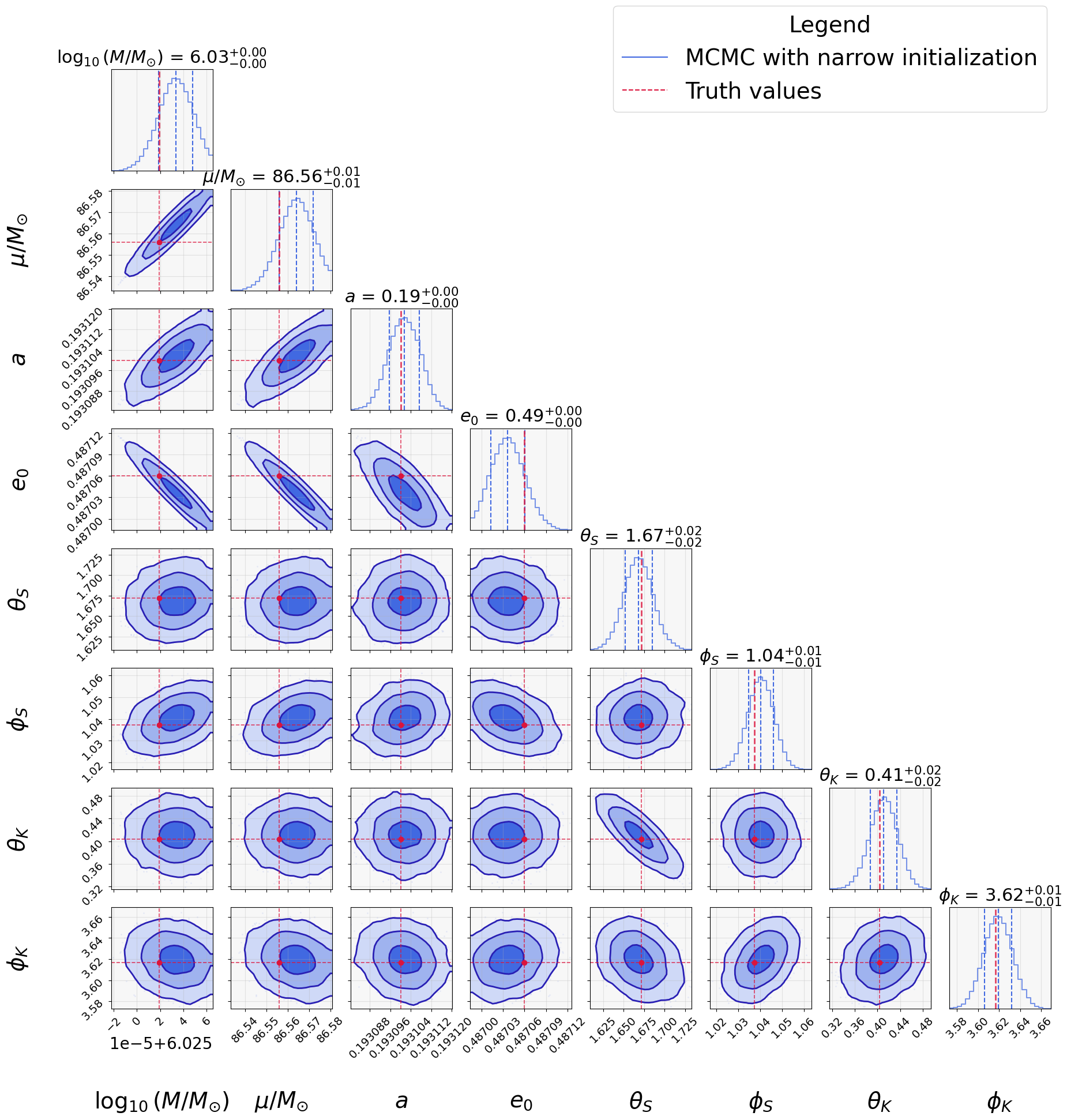}
  \caption{MCMC Parameter Recovery with Narrow Initialization.
Posterior distributions from MCMC chains initialized near true EMRI parameters (red lines) using setting (1). In the figure, the blue line represents the posterior probability distribution of
the EMRI parameters. 
The three contours represent the 1$\sigma$, 2$\sigma$, and 3$\sigma$ confidence regions of the posterior distribution.
Validates MCMC reliability with proper initialization.
  }
  \label{fig:near}
\end{figure*}

\begin{figure}[htb]
  \setlength{\arrayrulewidth}{0.2pt} 
  \centering
  \includegraphics[width=0.65\textwidth]{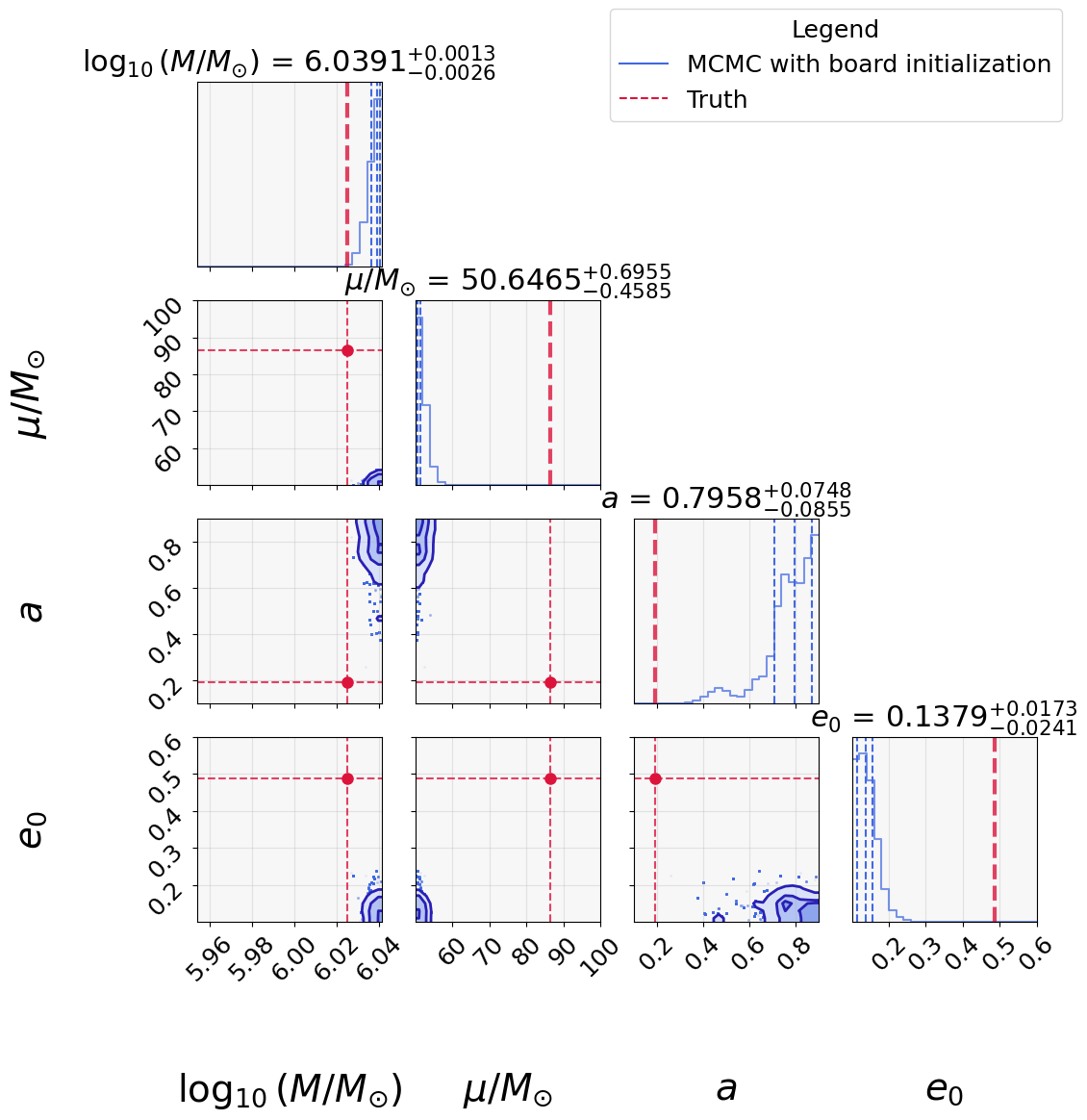}
  \caption{MCMC Parameter Estimation with Broad Initialization. Posterior distributions from MCMC chains initialized with wide priors.  
  In the figure, the blue line represents the posterior probability distribution of the EMRI parameters. 
  The dashed red lines indicate the true values of the parameters.
  The three contours represent the 1$\sigma$, 2$\sigma$, and 3$\sigma$ confidence regions of the posterior distribution.
}
  \label{fig:MCMC_nonear}
\end{figure}

\subsection{Result of Parallel Tempering Markov Chain Monte Carlo}\label{ptmcmc}

Running MCMC under setting (1) where all parameters are initialized  close  to the true values, we successfully recover the posterior distributions of the source parameters, see Figure~\ref{fig:near}.
This confirms that MCMC can accurately reconstruct EMRI parameters when starting from favorable initial conditions near the global maximum of the likelihood.
However, in realistic data analysis, the true parameters of an EMRI are unknown, and such informed initialization is not feasible.
To assess the practical performance of the MCMC sampler, we perform inference under setting (2), sampling intrinsic parameters drawn from broad priors. 
The resulting posteriors exhibit significant biases across all intrinsic parameters, with the true values lying outside the 3 $\sigma$ credible regions, see Figure~\ref{fig:MCMC_nonear}.
This behavior is due to the inherent multimodality and strong degeneracies in the EMRI data analysis, which cause MCMC chains to become trapped in local maxima. 
As a result, the sampler oscillates around suboptimal solutions without achieving convergence to the global optimum.
Therefore, under current settings, using PTMCMC alone is insufficient for robust scientific analysis of EMRI data, as it lacks the efficiency and global exploration capability required for such complex parameter spaces.
To the best of our knowledge, no existing study has demonstrated that MCMC methods can reliably recover the correct posterior distribution for EMRIs when initialized from broad and realistic priors.
This limitation highlights the severe challenges posed by the high-dimensional, multimodal, and degenerate structure of the EMRI likelihood surface.

\subsection{Result of Continuous Normalizing Flows}

\begin{figure}[htb]
  \setlength{\arrayrulewidth}{0.2pt} 
  \centering
  \includegraphics[width=0.55\textwidth]{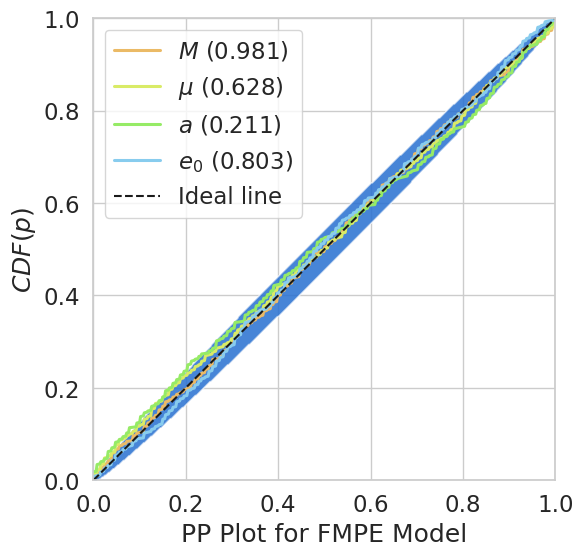}
\caption{{P-P plot for FMPE model, based on 500 simulated EMRI signals. The dashed line represents the ideal case, while the colored lines represent the empirical CDFs of the parameters. The shaded blue region indicates the 95\% confidence bands. All parameters lie well within the confidence region, demonstrating that the CNF captures posterior uncertainty with high fidelity. }}
  \label{fig:pp}
\end{figure}

To evaluate the performance of our Continuous Normalizing Flow (CNF) model (named Flow Matching Posterior Estimation, FMPE for short~\cite{Dax2023FlowMF, lipman2022flow}), we first tested it on 1,000 EMRI signals with source parameters randomly drawn from the priors defined in Table~\ref{table:priors}.
The resulting posterior samples were then used to generate the P-P (Probability-Probability) plot, as shown in Figure~\ref{fig:pp}.
The P-P plot shows that the cumulative distribution functions (CDFs) of the posterior percentile values for the intrinsic parameters closely align with the diagonal line.
This validates the statistical reliability and unbiasedness of our CNF-based posterior estimation across the full extent of the prior volume.
\begin{figure}[htb]
  \setlength{\arrayrulewidth}{0.2pt} 
  \centering
  \includegraphics[width=0.65\textwidth]{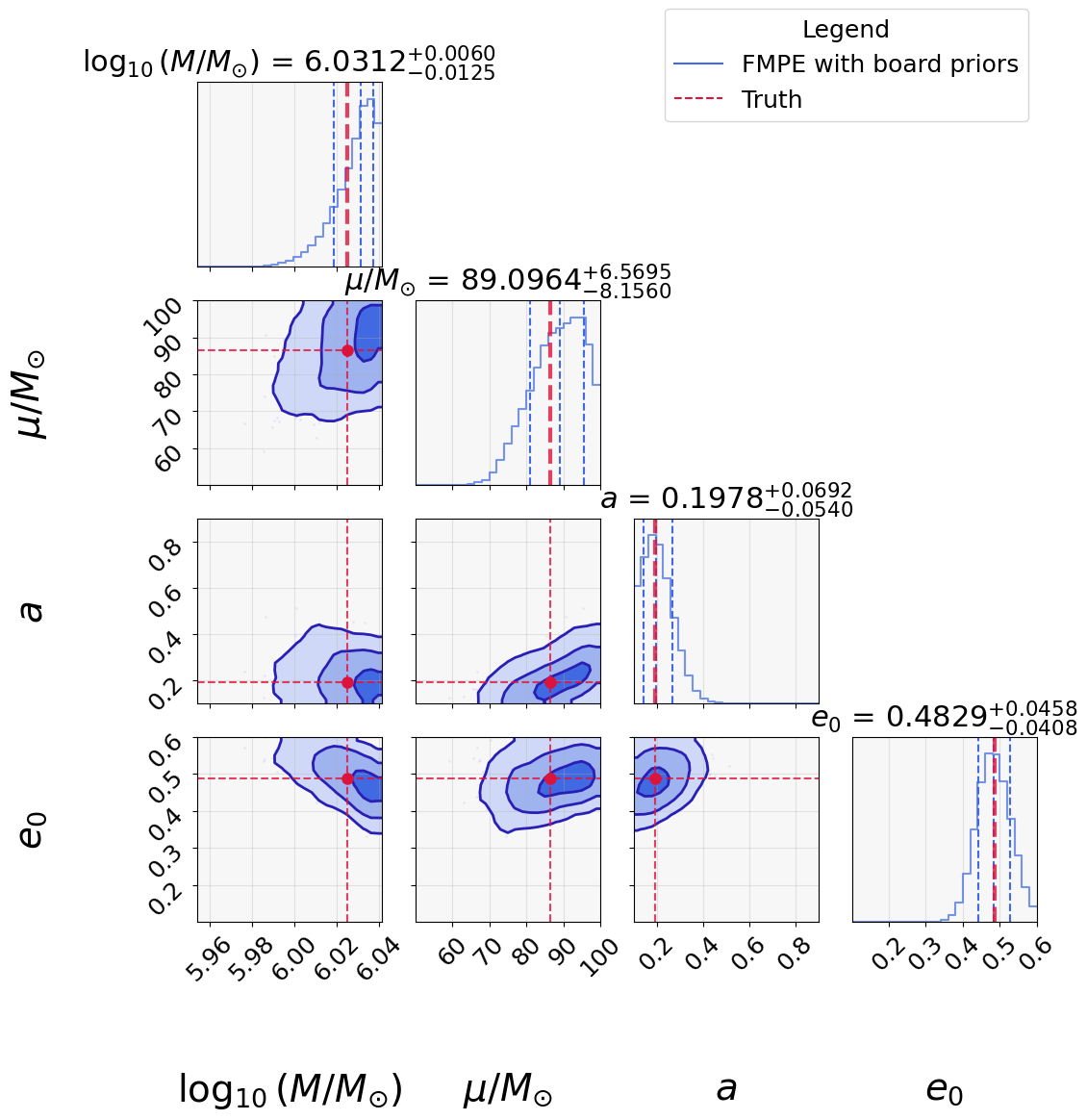}
  \caption{EMRI parameter estimation using FMPE under broad priors.
  Posterior distributions from FMPE initialized with wide priors. The blue shading represents the posterior probability distribution of EMRI parameters. Red dashed lines indicate injected true parameter values.
   The three contours represent the 1$\sigma$, 2$\sigma$, and 3$\sigma$ confidence regions of the posterior distribution.
}
  \label{fig:FMPE}
\end{figure}

We further compared the performance of FMPE and MCMC under the more realistic setting (2), in which intrinsic parameters are initialized from broad priors.
As shown in Figure ~\ref{fig:FMPE}, FMPE successfully recovered the posterior distributions of intrinsic parameters (\emph{e.g.} $M$ and $a$) with high accuracy, while clearly identifying the true values (red lines).
In stark contrast, MCMC under the same broad initialization (Figure~\ref{fig:MCMC_nonear}) failed to converge to the global maximum of the likelihood due to parameter degeneracies and local maxima.

Notably, FMPE achieves accurate posterior inference in minutes per inference, an improvement of several orders of magnitude over MCMC, which requires days to reach thermalization. This remarkable computational efficiency demonstrates that our pure machine learning-based approach enables rapid and unbiased parameter estimation for EMRIs, even in the challenging noise-dominated situations. 
Moreover, the inferred posterior distributions significantly narrow the initial Bayesian prior ranges. This prior refinement can inform and improve subsequent analyses for both machine learning models and traditional MCMC pipelines.
Especially, as shown in the next subsection, the high-fidelity posteriors generated by our FMPE model provide an ideal starting distribution for MCMC samplers. By initializing chains within the high-probability region of parameter space, FMPE enables a ``hot start'' for MCMC sampling that mitigates the risk of trapping in local maxima, addressing the major bottleneck in traditional EMRI parameter estimation.
\begin{table}[ht!]
    \centering
    \footnotesize
    \caption{{The table compares the parameters injected with those recovered by FM-MCMC and Eryn. The recovered values correspond to the median estimates, and the $1 \sigma$ confidence intervals are shown for each parameter. }} 
    \label{table:compare}
    \begin{tabular*}{\columnwidth}{@{\extracolsep{\fill}} l c @{\hspace{0.4em}} l c @{}}
        \toprule
        \hline
        \textbf{Parameter} &\ \ \ \  \textbf{Injected value} &\ \ \ \  \textbf{FM-MCMC} & \textbf{Eryn} 
         \\
        \midrule
        $\log_{10}{M} [M_\odot]$  & 6.0250 &  $6.0266_{-0.0024}^{+0.0020}$ & $6.0391_{-0.0026}^{+0.0013}$\\
        $\mu [M_\odot]$  & 86.5560 &  $86.7030_{-4.8716}^{+0.0525}$ & $50.6465_{-0.4585}^{+0.6955}$ \\
        $a$ & 0.19310 &  $0.1894_{-0.0000}^{+0.0172}$ & $0.7958_{-0.0855}^{+0.0748}$ \\
        $e_0$  & 0.48706 &  $0.4821_{-0.0195}^{+0.0062}$ & $0.1379_{-0.0241}^{+0.0173}$ \\
        \hline
        \bottomrule
    \end{tabular*}
\end{table}


\subsection{Result of FM-MCMC}
Building on the flow-based acceleration techniques, we propose in this work the pioneering FM-MCMC framework: a novel framework that seamlessly fuses the posterior sampling dynamics of CNF with the chain and temperature dynamics of PTMCMC.
\begin{figure}[htb]
  \setlength{\arrayrulewidth}{0.2pt} 
  \centering
  \includegraphics[width=0.65\textwidth]{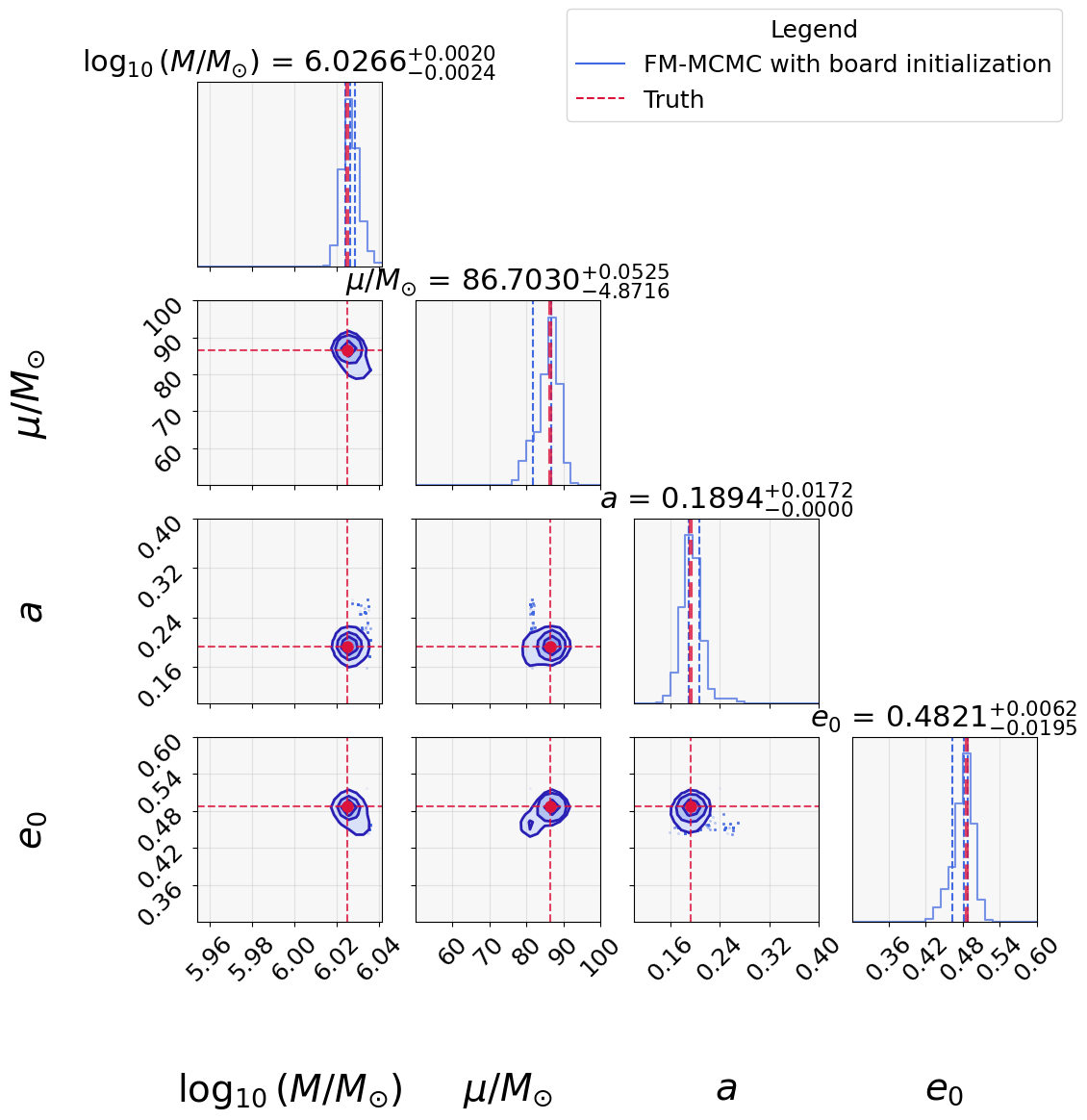}
  \caption{{The posterior distributions from the FM-MCMC framework initialized with wide priors. The blue contours show the posterior probability distribution of the EMRI parameters, while the red dashed lines indicate the true parameter values. The contours represent the 1$\sigma$, 2$\sigma$, and 3$\sigma$ confidence regions of the posterior distribution, highlighting the presence of multiple modes in the likelihood surface. The walkers shown in the figure correspond to the 0th walker, which has the largest negative log-likelihood among the 20 walkers.}
}
  \label{fig:flowmc}
\end{figure}

Figure~\ref{fig:flowmc} demonstrates the parameter recovery of FM-MCMC for the same EMRI  signal analyzed in the previous subsection under the broad priors listed in Table~\ref{table:priors}.
The full inference was completed within 48 hours on a single NVIDIA RTX 4090 GPU, demonstrating the feasibiliby of FM-MCMC for realistic EMRI data analysis tasks.
Crucially, unlike the case for standalone MCMC (Figure~\ref{fig:MCMC_nonear}), FM-MCMC converges unambiguously to the true values (red lines in Figure \ref{fig:flowmc}) across all intrinsic parameters.
Moreover, as quantified in Figure~\ref{fig:FMPE} and Figure~\ref{fig:flowmc}, FM-MCMC also achieves significant improvements over FMPE. 
This enhanced performance arises from the synergistic integration of the rapid exploration of machine learning with the refined sampling of MCMC, combining the speed of machine learning with the statistical reliability of Bayesian sampling.
{
We note that while normalizing flows are sometimes reported to under-represent small modes, in our framework the CNF does not serve as the final posterior estimator. Instead, CNF proposals are used to accelerate initialization, and the subsequent PTMCMC stage ensures more precise Bayesian inference. As shown in Figure~\ref{fig:flowmc} and in the additional walk results in Appendix~\ref{A2}, FM-MCMC successfully identifies and explores multiple modes in the likelihood surface, including subdominant ones. This confirms that the framework does not collapse to a single dominant mode but robustly preserves the multimodal structure of the EMRI parameter space.  }

Clear evidence presented in Table~\ref{table:compare} offers direct quantitative comparisons between the injected signal parameters and those recovered by FM-MCMC and standalone MCMC. 
This rigorously confirms the unprecedented precision of the FM-MCMC framework, which, to the best of our knowledge, marks the first successful estimation of the intrinsic parameters for Kerr EMRIs under realistic noise conditions of LISA-like missions with fully unrestricted priors.
This advancement finally brings within reach a deep understanding of the astrophysical formation channels ~\cite{PhysRevX.6.041015, Abbott_2016}, unprecedented precision tests of GR including direct validation of the ``no hair'' theorem~\cite{Gair_2013, PhysRevLett.130.241402}, and a novel pathway for probing potential dark matter signatures around massive black holes~\cite{PhysRevX.6.041015, PhysRevD.94.084002, PhysRevLett.116.221101} with EMRI observations.


\section{Conclusion}\label{sec:Conclusion}

This work establishes FM-MCMC, the first machine learning-enhanced Bayesian framework integrating CNFs with PTMCMC, which resolves the long-standing challenge in EMRI data analysis, that of global convergence in high-dimensional, multimodal parameter spaces. 
Conventional MCMC methods are prone to become trapped in local likelihood maxima, leading to systematic biases in the recovery of intrinsic EMRI parameters under realistic instrumental noise conditions and initial priors.
In contrast, given flow matching to learn and deploy globally informed proposal distributions, that by dynamically generating high-likelihood regions via CNFs and refining them through PTMCMC, FM-MCMC enables robust exploration of degenerate, high-dimensional parameter spaces, while achieving order-of-magnitude improvements in computational efficiency and, more importantly, ensuring statistically reliable inference.
{
For EMRI analysis, FM-MCMC alleviates parameter-estimation degeneracies arising from both instrumental noise and the complex harmonic structure of AAK waveforms with higher modes.
Our results demonstrate that FM-MCMC can successfully recover EMRI parameters even under broad prior assumptions, highlighting its robustness compared to traditional methods.
Looking forward, the extension of this framework to realistic data conditions, including uncertain noise PSDs, astrophysical foregrounds, and potential gravitational-wave backgrounds, will be an essential step.
Ultimately, such developments will make EMRI parameter estimation a feasible and powerful tool for unlocking the rich scientific opportunities enabled by future space-based gravitational-wave observations.}

At last but not least, FM-MCMC enables real-time analysis of EMRI signals, reducing inference times from days to hours on a single GPU, and establishing a scalable pipeline for the future planned space-based antennas like Taiji, TianQin and LISA.
Future work will extend this hybrid paradigm to multi-source inferences and continue to advance gravitational-wave astronomy into the era of intelligent data exploitation.

\section{Methodological Framework}\label{Sec:model}

\begin{figure*}[ht!]
  \centering
  \includegraphics[width=1.0\textwidth]{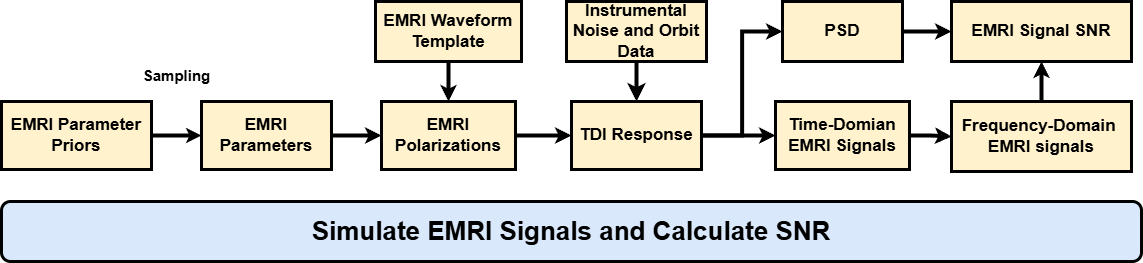}
  \caption{
  The framework for EMRI signal generation and pre-processing pipeline. This framework begins with parameter priors and waveform generation, followed by orbital data and instrumental noise input, TDI response calculation, noise spectral analysis, and time-to-frequency domain transformation through pre-processing steps. 
  }
  \label{fig:pre-processing}
\end{figure*}

The construction of the datasets for EMRI analysis in this work rests on four key components: 
(1) the numerical synthesis of EMRI waveforms, 
(2) specification of the detector orbit configuration, 
(3) application of the time-delay interferometry (TDI), 
and (4) pre-processing of data for model training. 
As illustrated in Figure~\ref{fig:pre-processing}, this methodological framework sequentially integrates these components to ensure physical fidelity and computational tractability. 
The following subsections provide detailed descriptions of each operational stage within this pipeline.

\subsection{Extreme mass ratio inspiral waveforms}
The accurate waveform modeling for EMRI systems constitutes a critical challenge, requiring high-fidelity implementations of gravitational self-force (GSF) calculations and black hole perturbation theory to achieve the sub-radian phase precision for meaningful scientific inference \cite{Barack:2018yvs,Pound:2021qin}.
Perturbative expansions of the binary metric have been applied at the first order for solving general orbits around Kerr black holes~\cite{vandeMeent:2017bcc}, with significant progress also made in second-order calculations~\cite{Pound:2019lzj,Warburton:2021kwk}. 
However, the computational efficiency of numerical gravitational self-force calculations remains insufficient to meet the demands of data analysis. 
Recent developments in effective-one-body models~\cite{Albertini_2024, Shen_2023} synthesize GSF data into resummed potentials and try to balance between physical self-consistency and computational efficiency.
On the other hand, rapid generation of approximate kludge models has been developed~\cite{Barack:2003fp,Babak:2006uv,Chua:2015mua,Chua:2017ujo}.
Early analytic kludge (AK) models~\cite{Barack:2003fp}, while sacrificing some accuracy, offer significantly improved computational efficiency. 
Numerical kludge (NK) models~\cite{Gair:2005ih,Babak:2006uv}, on the other hand, achieve high-fidelity EMRI waveforms by fitting the perturbative calculations. The augmented analytic kludge (AAK) models~\cite{Chua:2015mua,Chua:2017ujo,Chua:2020stf} incorporate the NK model, enabling the generation of high-precision EMRI waveforms without significantly increasing computational costs. 
To date, the most advanced computational framework \texttt{FastEMRIWaveforms} (\texttt{FEW})~\cite{Katz:2021yft,Chua:2020stf} can rapidly compute EMRI waveforms in the time domain.  For data preparation, we employ the AAK model implemented in \texttt{FEW} to generate the EMRI waveforms. The model’s sub-radian phase accuracy satisfies the precision criteria required for reliable parameter inference in this study. 

\subsection{Detector orbit configuration}
Both the Taiji and LISA antennas consist of three spacecrafts (S/Cs) forming near-equilateral triangular constellations in heliocentric orbits. 
By continuously monitoring variations in the optical path lengths between S/Cs using laser interferometers, these detectors can measure the imprints induced by incident GWs.
In this paper, we model the single-arm GW response  in terms of laser frequency modulations
as defined in ~\cite{Katz:2022yqe} and  ~\footnote{\url{https://lisa-ldc.lal.in2p3.fr/static/data/pdf/LDC-manual-002.pdf}},
which explicitly depends on the detector's orbit  configuration. 
As mentioned, we take Taiji as the representative of LISA-like missions, and employ an equal-arm analytic model to describe its orbit, where the coordinates of S/C$_i$ ($i \in \{1,2,3\}$) in the Solar System Barycentric (SSB) frame read~\cite{LDC_Radler_manual_v2,Du2023AdvancingSG}: 
\begin{align}  
    \bm{R}_{i}(t) &= \bigg\{ R\cos\beta + \frac{L}{2\sqrt{3}}\left[\frac{1}{2}\sin2\beta\sin\gamma_i  
     - (1+\sin^2\beta)\cos\gamma_i\right], \nonumber \\  
     & \quad \quad  R\sin\beta + \frac{L}{2\sqrt{3}}\left[\frac{1}{2}\sin\beta\cos\gamma_i  
     - (1+\cos^2\beta)\sin\gamma_i\right], \nonumber \\  
     & \quad \quad  -\frac{L}{2}\cos\left[\beta - \gamma_i\right] \bigg\}, 
\end{align}  
where 
\begin{align}
    \beta(t) \equiv \frac{2\pi }{T}t + \beta_0, \quad 
    \gamma_i \equiv \frac{2\pi (i  - 1)}{3} + \gamma_0.   
\end{align}
Related parameters are specified as follows: 
the nominal arm length of Taiji is \( L = 3 \times 10^9 \, \mathrm{m} \), 
the orbital radius and period of Taiji's guiding center are
\( R = 1 \, \mathrm{AU} \), \( T = 1 \, \mathrm{yr} \), respectively, and  the initial phase angles are set to \( \beta_0 = \gamma_0 = 0 \),  without loss of generality.
{
Taiji has a nominal arm length of 3 million kilometers (compared to LISA’s 2.5 million kilometers) and a readout noise budget of 8 ${\rm pm}/\sqrt{\rm Hz}$ (versus LISA’s 15 ${\rm pm}/\sqrt{\rm Hz}$). 
Thus Taiji is expected to have slightly better sensitivity to GW signals (\emph{e.g.} approximately 1.3 times better at 1 mHz), see Ref.~\cite{Luo:2021qji} for the overall design of Taiji, and Figure~\ref{fig:TD} of Ref.~\cite{hewangsc} for the comparison among Taiji, LISA, and TianQin's sensitivities.
As a result, Taiji may detect a slightly larger number of observable EMRIs, which in turn would lead to greater demand for the analysis of these signals. 
However, due to the current significant uncertainties in EMRI population models, it is not yet feasible to provide a quantitative comparison.
}

\subsection{Time-delay interferometry}
{For space-based GW detection, 
the inequality of arm and the instability of lasers causes the laser frequency to be orders of magnitudes larger than the target GW signals~\cite{Tinto_time_2021}. 
To  achieve the designed   sensitivity, time-delay interferometry (TDI) is proposed effectively to suppress laser frequency noise in the data processing stage~\cite{Armstrong_time_1999,wtn_tdi}.}
The basic principle of TDI is to apply appropriate time delays to the single-arm measurements and combine them to synthesize an effective equal-arm interferometer. 
{First-generation TDI combinations perform well for static unequal-arm configurations, while second-generation TDI further extends this capability under situations with  and time-varying arm lengths, hence is more suitable for the realistic Taiji and LISA orbits~\cite{Tinto:2022zmf}.}
For example, the second-generation Michelson TDI combination  \(X(t)\) is defined as~\cite{Otto:2015erp}:
\begin{align}
X(t) =& \ y_{1'} + y_{3,2'} + y_{1,22'} + y_{2',322'}  \nonumber \\  
&+ y_{1,3'322'} + y_{2',33'322'} + y_{1',3'33'322'}  \nonumber \\ 
&+ y_{3,2'3'33'322'}  - y_{1} - y_{2',3} - y_{1',3'3}   \nonumber \\ 
&- y_{3,2'3'3} - y_{1',22'3'3}  - y_{3,2'22'3'3}  \nonumber \\ 
&- y_{1,22'22'3'3} - y_{2',322'22'3'3}.
\end{align}  
where $y$ denotes the single-arm measurement, and the digits following the comma represent time-delay operations applied according to the corresponding arm lengths (see Ref.~\cite{Otto:2015erp} for details). 
The \(Y\) and \(Z\) channels are obtained by cyclically permuting the indices according to the rule $1 \rightarrow 2$, $2 \rightarrow 3$, and $3 \rightarrow 1$. 
We further derive the widely adopted quasi-noise-orthogonal $\{A,E,T\}$ channels from  $\{X,Y,Z\}$ as follows: 
\begin{align}
A &= \frac{1}{\sqrt{2}} (Z - X), \nonumber \\ 
E &= \frac{1}{\sqrt{6}} (X - 2Y + Z), \nonumber \\
T &= \frac{1}{\sqrt{3}} (X + Y + Z).
\end{align}  

{
Since data analysis is performed in the form of TDI variables, 
it would be  beneficial to provide an explanation on the impacts of detector orbit and TDI  on the EMRI waveforms. 
Firstly, the rotation of the detector constellation causes time-varying amplitude modulation.  
Secondly, the translation motion of the detector induces a Doppler frequency shift.
Thirdly, TDI combination involves the summation of waveforms at different times. 
For the impact of second-generation TDI, an intuitive illustration to the effect of TDI can be given  under the equal-arm approximation.
Taking the second-generation Michelson-$X$ channel as an example, the Fourier transform of TDI result can be expressed as 
\begin{equation}
    \Tilde{X}(f) \approx - 4 e^{-i3u} \sin 2u \sin u \tilde{X}_0(f),
\end{equation}
with $u \equiv 2\pi f L / c$, $\tilde{X}_0(f)$ being the response of a  Michelson interferometer,  and the preceding factor represents the additional correction  introduced by the TDI combination (relative to $\tilde{X}_0(f)$, also see the result for the 1st-generation in Ref.~\cite{LDC_Radler_manual_v2}). 
The GW responses in the $A$ and $E$ channels are equivalent to two Michelson interferometers oriented at 45 degrees relative to each other~\cite{Wang:2020fwa}, whereas the $T$ channel is insensitive to signals and hence referred to as the ``noise'' channel or ``null'' channel.
}

{
In this study, to leverage their complementary sensitivity, we utilize both the $A$ and $E$ channels for the likelihood calculation in the final MCMC parameter estimation. 
For the training of the continuous normalizing flow model (described in Section 4.6), however, we used data exclusively from the $A$ channel. This choice was made for computational efficiency, as the $A$ channel alone proved sufficient to generate an effective proposal distribution for initializing the MCMC sampling. 
The $T$ channel, which is typically used for noise characterization, was not included as this study operates under the assumption of known noise PSDs, also see the Discussion section. 
The TDI response calculations for all channels were implemented using the open-source tool \texttt{FastLISAResponse}~\footnote{\url{https://github.com/mikekatz04/lisa-on-gpu/tree/master}}. Each data sample (used for training, as detailed in Section 4.5) comprises two-month-duration TDI-$A$ channel measurements, simulated at a sampling interval of 25 seconds. The GPU-accelerated heterogeneous computing architecture employed for both waveform generation and response emulation~\cite{Katz_2022} enables the generation of single waveforms with latency under one second, while maintaining the required numerical precision~\cite{burke2024accuracyrequirementsassessingimportance}.
}

\subsection{Data pre-processing}
Although GPU acceleration significantly improves the efficiency of waveform generation, the high dimensionality of the raw time-domain data still poses difficulties for machine learning models.
Frequency domain analysis~\cite{bayle2022overview,speri2022roadmap,burke2024accuracyrequirementsassessingimportance} has proven to be an exceptionally effective method for estimating parameters in EMRI analysis. 
Therefore, we transform the time-domain samples to the frequency domain using the Fast Fourier Transform (FFT) algorithm.

For discrete time-series $s_{\rm EMRI}^A(t_n)$ with $N$ samples, the spectral components can be derived as
\begin{equation}
\begin{aligned}
    S_{\rm EMRI}^A(f_k) = \mathcal{F}\{w(t_n) \cdot s_{\rm EMRI}^A(t_n)\} =  \\
    \sum_{n=0}^{N-1} w(t_n) s_{\rm EMRI}^A(t_n) e^{-i2\pi kn/N},
\end{aligned}
\end{equation}
where $t_n = n\Delta t$ defines the sampling times and $\Delta t$ the sampling interval, $f_k = k/(N\Delta t)$ denotes frequency bins, and $w(t_n)$ is the Tukey window ($\alpha=0.05$).  $\mathcal{F}$ denotes the  FFT algorithm.
The windowed signal satisfies
\begin{equation}
s_{\rm EMRI}^{\rm window}(t_n) = \begin{cases} 
w(t_n) \cdot s_{\rm EMRI}^A(t_n), & 0 \leq n < N_{\rm orig} \\
0, &N_{\rm orig} \leq n < N_{\rm pad} ,
\end{cases}
\end{equation}
with $N_{\rm orig}$ being the original signal length and $N_{\rm pad}$ the zero-padded length.

In the EMRI analysis, conventional min-max normalization may induce information degradation. Therefore, we propose a pre-processing scheme that combines spectral whitening with standardization.
For the TDI-$A$ channel, we first compute its noise power spectral density (PSD) in terms of the test-mass acceleration noise $S_{\rm acc}(f)$ and the optical metrology system noise
$S_{\rm oms}(f)$, 
\begin{equation}
\begin{aligned}
\text{PSD}(f) &= 32 \sin^2(4u) \sin^2(2u) \Big[  S_{\rm oms}(f)\big(2 + \cos (2u)\big)\\
&\quad + 2S_{\rm acc}(f)\big(3 + 2\cos (2u) + \cos (4u) \big)  \Big],
\end{aligned}
\end{equation}
where $u \equiv \pi Lf/c$ is the normalized frequency, and $S_{\rm oms}(f)$, $S_{\rm acc}(f)$ take the designed  spectral profiles of Taiji (see \emph{e.g.} Ref.~\cite{Wang:2020vkg}).  
The amplitude spectral density (ASD) is then derived as $\text{ASD} \equiv  \sqrt{\text{PSD}}$. 
To ensure that the SNR remains consistent before and after whitening, we introduce a scale factor:
\begin{equation}
\kappa = \sqrt{\frac{N_t}{4\Delta t}},
\end{equation}
with $N_t$ being the number of time samples. The whitening process is ultimately implemented through spectral normalization:
\begin{equation}
\tilde{s}_{\rm whitened}(f) = \frac{s_{\rm EMRI}^{\rm window}(t_n)}{\text{ASD} \cdot \kappa} ,
\end{equation}
This pipeline effectively decouples noise correlations while maintaining the phase coherence and amplitude characteristics of EMRI GW signals.

We then establish a physically consistent noise injection framework based on whitened signals. The complex Gaussian noise in the frequency domain is modeled with independent real and imaginary components,
\begin{equation}
n(f) = n_{\rm real}(f) + i \cdot n_{\rm imag}(f), \quad n_{\rm real},n_{\rm imag} \sim \mathcal{N}(0,\sigma^2),\label{eq10}
\end{equation}
where the variance $\sigma^2=1$ is guaranteed by whitening normalization. The noise injection is implemented through spectral superposition,  
\begin{equation}
s'(f) = \tilde{s}_{\rm whitened}(f) + n(f).\label{eq11}
\end{equation}
This process preserves the phase coherence of the signals while ensuring consistency with Taiji’s noise PSD characteristics. This completes the full pipeline for generating individual EMRI gravitational wave signals.

\subsection{Dataset generation}\label{datagen}
EMRI data are inherently challenging to process due to their large size~\cite{Babak_2015}, and this difficulty is further amplified in machine learning applications, where models are trained on batched data.
After the FFT,  each sample can reach lengths of up to millions.
While dimensionality reduction techniques like image-based downsampling or adaptive pooling could theoretically compress the data volume, these methods risk losing vital relativistic features present in EMRI waveforms.
To preserve the full harmonic coherence and orbital features essential for robust parameter inference, the TDI-$A$ channel EMRI signals were kept in their pristine, noise-free form within the training dataset. 
During each training iteration, distinct realizations of instrumental noise were dynamically generated and injected into the clean signal data, which accurately emulates the stochastic noise characteristics encountered in real observations.
Specifically, we first perform random sampling within the prior parameter ranges listed in Table ~\ref{table:priors}, selecting 20,000 EMRI signal parameters with SNRs exceeding 60 to form the training dataset. 
An additional 2,000 parameter sets meeting the same SNR criterion were selected to constitute the test set.

\subsection{Accelerating MCMC sampling with Continuous Normalizing Flows}\label{fmmc}


This study introduces an innovative method, FM-MCMC, which accelerates MCMC sampling through CNFs. By integrating CNFs into the widely used \texttt{Eryn}~\cite{Karnesis_2023} MCMC framework  for space-based GW data analysis, the proposed approach significantly enhances the efficiency of parameter estimation.
Specifically, we first train a CNF model to rapidly generate posterior distributions for EMRI's intrinsic parameters. Through a dynamic matching mechanism, the sample size generated by the model automatically adapts to \texttt{Eryn} MCMC's walker count and temperature ladder configuration. Simultaneously, based on the posterior distribution characteristics output by CNFs, the system intelligently optimizes \texttt{Eryn} MCMC's initial parameter space to achieve automated contraction of prior distribution ranges.

\subsubsection{Continuous Normalizing Flows}\label{cnf}

While current machine learning applications in natural sciences frequently employ Neural Posterior Estimation (NPE)~\cite{papamakarios2018fastepsilonfreeinferencesimulation, Du2023AdvancingSG, liang2024rapidparameterestimationmerging, Dax_2021, gebhard2023inferringatmosphericpropertiesexoplanets} with Discrete Normalizing Flows (DNFs)~\cite{rezende2015variational, JMLR:v22:19-1028}, recent advances suggest that flow matching with CNFs~\cite{chen2019neuralordinarydifferentialequations} – termed FMPE~\cite{Dax2023FlowMF, lipman2022flow} – offers superior training efficiency and accuracy~\cite{liang2024rapidparameterestimationmerging, Dax2023FlowMF, gebhard2023inferringatmosphericpropertiesexoplanets}.
, we examine two probability distributions over $\mathbb{R}^d$ characterized by densities $q(\theta_0)$ (base distribution $q_0$) and $q(\theta_1|x)$ (target posterior $q_1$), where $x$ represents observational data. The fundamental objective of CNFs is to construct a diffeomorphism $f: \mathbb{R}^d \to \mathbb{R}^d$ satisfying the transport condition: if $\theta_0 \sim q_0$, then $f(\theta_0) \sim q_1$. 
CNFs realize this transformation via a continuous process governed by a temporal parameter $t \in [0,1]$, where the dynamics are defined by a neural velocity field $v_{t,x}:\mathbb{R}^d \to \mathbb{R}^d$,
\begin{equation}
\frac{d}{dt}\phi_{t,x}(\theta) = v_{t,x}(\phi_{t,x}(\theta)), \quad \phi_{0,x}(\theta) = \theta_0, \quad \phi_{1,x}(\theta) = \theta_1, \label{q1}
\end{equation}
where $\phi_{t,x}$ represents the flow map transporting samples from $q_0$ to $q_1$ under the conditioning variable $x$. In this context, $x$ corresponds to the EMRI signal in the presence of noise. Target distribution samples are generated by numerically integrating Eq.~\ref{q1} with initial conditions drawn from $q_0$.
The associated probability density evolution follows the continuity equation,
\begin{equation}
\frac{\partial}{\partial t} q_{t,x}(\theta) + \nabla_\theta \cdot (q_{t,x}(\theta) v_{t,x}(\theta)) = 0,
\end{equation}
yielding the instantaneous density via 
\begin{equation}
q(\theta|x) = q_0(\theta_0) \exp\left(-\int_0^1 \nabla_\theta \cdot v_{t,x}(\theta_t) dt\right). \label{q2}
\end{equation}
This formulation enables simultaneous density evaluation and sampling through adaptive ODE solvers.

Following flow matching principles, we parameterize the velocity field through regression on conditional vector fields,
\begin{equation}
u_t(\theta|\theta_1) = \theta_1 - \theta,
\end{equation}
which induce Gaussian probability paths
\begin{equation}
p_t(\theta|\theta_1) = \mathcal{N}\left(\theta_1 t, (1-t)^2 I_d\right).
\end{equation}
These paths interpolate between a standard normal distribution at $t=0$ and a posterior distribution centered at $\theta_1$ as $t\to1$. The training objective is to minimize the expected deviation between learned and target vector fields,
\begin{equation}
\mathcal{L}_{\mathrm{FM}} = \mathbb{E}_{t \sim \mathcal{U}(0,1)} \mathbb{E}_{\theta_1 \sim p(\theta)} \mathbb{E}_{x \sim p(x|\theta_1)} \mathbb{E}_{\theta_t \sim p_t(\theta_t|\theta_1)} \| r \|^2,
\end{equation}
\begin{equation}
r = v_{t,x}(\theta_t) - u_t(\theta_t|\theta_1),
\end{equation}
where $v_{t,x}$ learns to transport samples while preserving density consistency through the divergence term in Eq.~\ref{q2}.

\subsubsection{High-precision Bayesian inference}

Our framework employs CNFs, trained through flow matching, to enable rapid posterior estimation for EMRI. The trained model generates \>$10^3$ posterior samples in just a few minutes, providing high-quality, optimized initial proposals for PTMCMC sampling.
The CNF sample generation dynamically adapts to PTMCMC's chain configurations and temperature ladder parameters. Final posterior distributions are derived from thermally equilibrated PTMCMC chains.

Our training set comprises 20,000 clean EMRI waveform samples, each waveform representing a two-month Taiji observation window sampled at fixed intervals, totaling 200,000 data points. 
The neural architecture comprises two core components: (1) A 1D convolutional dimensionality reduction encoder that compresses raw GW signals from 200,000 to 2,048 dimensions through nonlinear operations while preserving critical phase information; (2) A CNF network with 21 residual blocks, achieving progressive feature compression from 3072 to 4 dimensions. 

This hierarchical architecture achieves synergistic optimization of EMRI signal embeddings \( x \), temporal evolution parameter \( t \), and dynamic parameter states \( \theta_t \) through a multi-phase feature interaction mechanism. By progressively integrating the three critical components, the framework establishes an accurate mapping to the vector field \( v_{t,x}(\theta_t) \). 
The training was conducted on an NVIDIA RTX 4090 GPU, utilizing the Adam optimizer (initial learning rate 0.0001) with a 2000 epoch cosine annealing schedule. The full training cycle required approximately 48 hours.

\section{Dissusion}
{
In this work, we have proposed and demonstrated the FM-MCMC framework for EMRI parameter estimation. Compared with traditional methods, FM-MCMC significantly improves computational efficiency while maintaining high accuracy, thereby making EMRI parameter estimation more tractable. Our results show that this approach can effectively overcome the difficulties associated with highly multimodal likelihood surfaces and efficiently recover the intrinsic parameters of EMRIs.}

{
While our current focus has been on demonstrating the feasibility and efficiency of FM-MCMC, realistic EMRI data analysis will face further challenges. In addition to instrumental noise, one must also consider uncertainties in the noise PSD, astrophysical foregrounds such as galactic white dwarf binaries, and gravitational wave backgrounds from other astrophysical sources. Developing methods that remain effective under such complex noise and confusion conditions will be a key direction of future research.}

{
We also note that our current framework, similar to standard MCMC methods, assumes that the signal parameters lie within the predefined prior distribution used to train the CNF model. Inference cannot be guaranteed if the data fall completely outside this defined parameter range, and the development of robust out-of-distribution handling, potentially using methods such as transfer learning or fine-tuning to adapt the model to new data regimes, remains an important direction for future work.
}

{
So far, the so-called ``global fitting'' problem--the joint analysis of all signals and noises--remains an active research area. 
Several prototype designs for global fit pipeline (\emph{e.g.} Refs.~\cite{Littenberg_2023,PhysRevD.111.024060,deng2025modularglobalfitpipelinelisa}) have been proposed, which generally adopt a modular structure. 
Namely, the overall workflow is divided into modules, such as the characterization of noise, foreground and background, as well as the parameter estimation of massive black hole binaries (MBHBs), Galactic binaries (GBs), and EMRIs, with exchange of information and residual data between them. 
Among these target signals,  
MBHB mergers are high-SNR (up to $\mathcal{O}(10^3)$ - $\mathcal{O}(10^4)$) transient signals, while GBs are persistent signals highly localized in the frequency domain. 
It is therefore natural to analyze them in the pipeline's early stages, and noise characterization modules should also be invoked for the construction of likelihoods.  
Besides, verification sources (such as verification GBs) and auxiliary data of the spacecrafts  (such as the readouts of gravitational reference sensors, fluctuations of temperature and residual magnetic fields near core payloads, platform attitude jitters, etc.) can also provide supplementary information for calibrating the noise profile. }
{
Consequently, it is reasonable to focus here on the analysis of EMRI, under the condition that other bright signals have been extracted and the spectra of noise/foreground/background are known in the prior stages. 
Crucially, the key challenge for EMRI parameter estimation, and the core contribution of this work, lies in overcoming local optima and computational complexity to enable inference across relatively wide priors. }
{
We show that compared with traditional methods, FM-MCMC can efficiently achieve global maxima under highly multimodal likelihood surfaces, making EMRI parameter estimation more tractable.
The proposed algorithm is designed to ultimately function as the ``EMRI parameter estimation module'' in the full global fitting pipeline. }
{
In future work, we will also explore the potential interaction of EMRI signals with instrumental noises and GW foreground/background,  upon integration into the full global fitting pipeline.
For example, the non-stationarity in instrumental noise and astrophysical foreground can be handled through methods such as  whitening in the time-frequency domain~\cite{PhysRevD.102.124038,gpmh-1hqx}, an approach fully compatible with our framework.
}

\section*{Acknowledgments}
This study is supported by the National Key Research and Development Program of China (Grant No. 2021YFC2201901, Grant No. 2021YFC2203004, Grant No. 2020YFC2200100 and Grant No. 2021YFC2201903). International Partnership Program of the Chinese Academy of Sciences, Grant No. 025GJHZ2023106GC. We also gratefully acknowledge the financial support from Brazilian agencies Funda\c{c}\~ao de Amparo \`a Pesquisa do Estado de S\~ao Paulo (FAPESP), 
Fundação de Amparo à Pesquisa do Estado do Rio Grande do Sul (FAPERGS), Funda\c{c}\~ao de Amparo \`a Pesquisa do Estado do Rio de Janeiro (FAPERJ), Conselho Nacional de Desenvolvimento Cient\'{\i}fico e Tecnol\'ogico (CNPq), and Coordena\c{c}\~ao de Aperfei\c{c}oamento de Pessoal de N\'ivel Superior (CAPES).
This work is supported by High-performance Computing Platform of Peking University.

\printbibliography
\appendix
\section{FM-MCMC Walk Results}
\begin{figure}[htb]
  \setlength{\arrayrulewidth}{0.2pt} 
  \centering
  \includegraphics[width=0.85\textwidth]{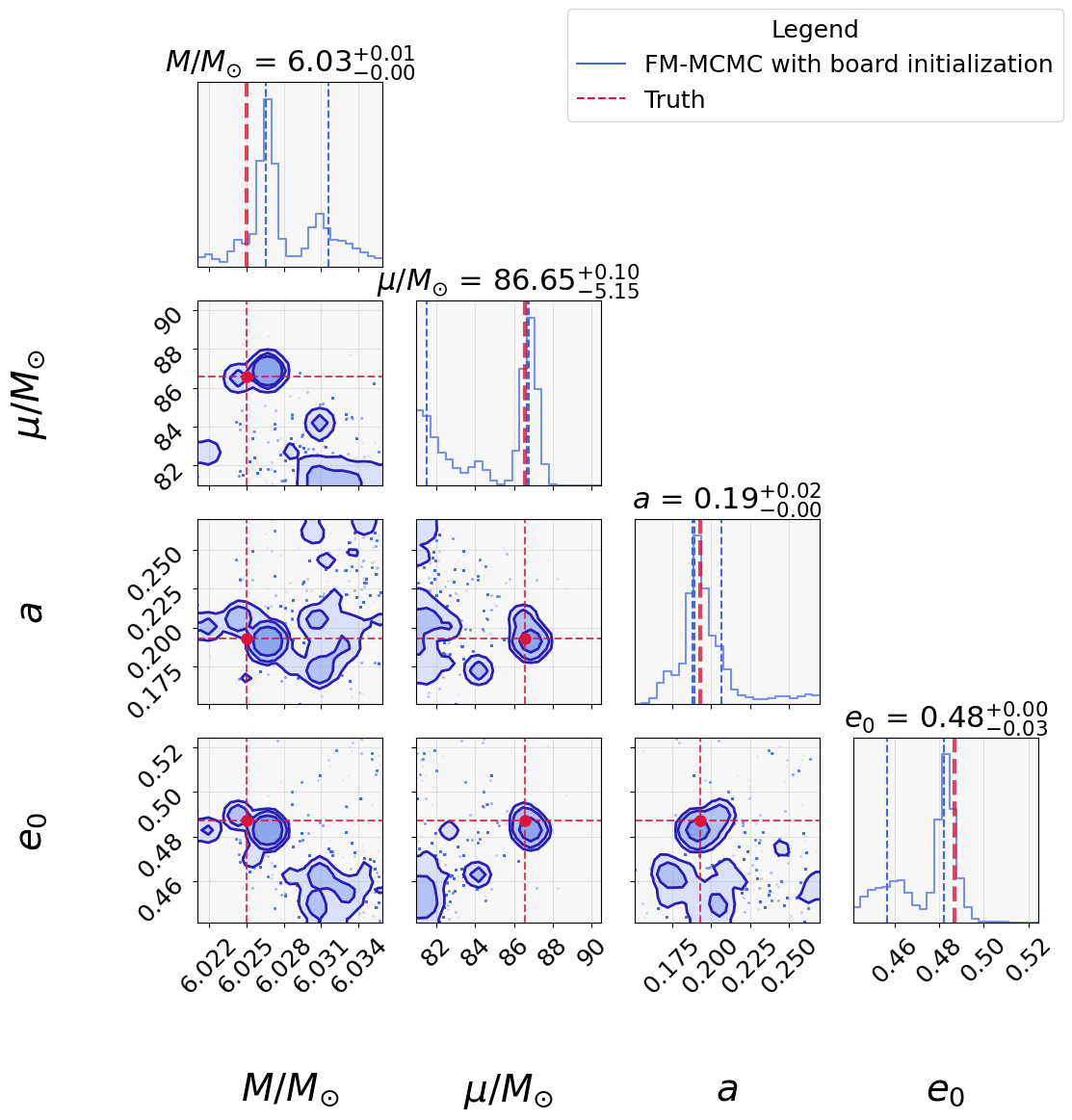}
  \caption{{This figure shows the posterior distribution for walkers 0 and 7 in the FM-MCMC run. The plot highlights the multimodal structure of the posterior distribution, with clear evidence that FM-MCMC is able to explore multiple modes of the likelihood surface. The walkers effectively navigate between different modes, demonstrating the method’s superior global exploration capabilities.}
}
\label{A1}
\end{figure}

{
In this appendix, we provide additional walk results for the FM-MCMC framework, showing the posterior distributions for walkers 0 and 7. As shown in Figure~\ref{A1}, these walkers clearly demonstrate the multimodal nature of the posterior distribution, revealing how FM-MCMC is able to explore multiple modes effectively. The walkers successfully navigate between different modes of the likelihood surface, providing a clear illustration of FM-MCMC’s ability to avoid local maxima and explore the global posterior distribution.}

\begin{figure}[htb]
  \setlength{\arrayrulewidth}{0.2pt} 
  \centering
  \includegraphics[width=0.85\textwidth]{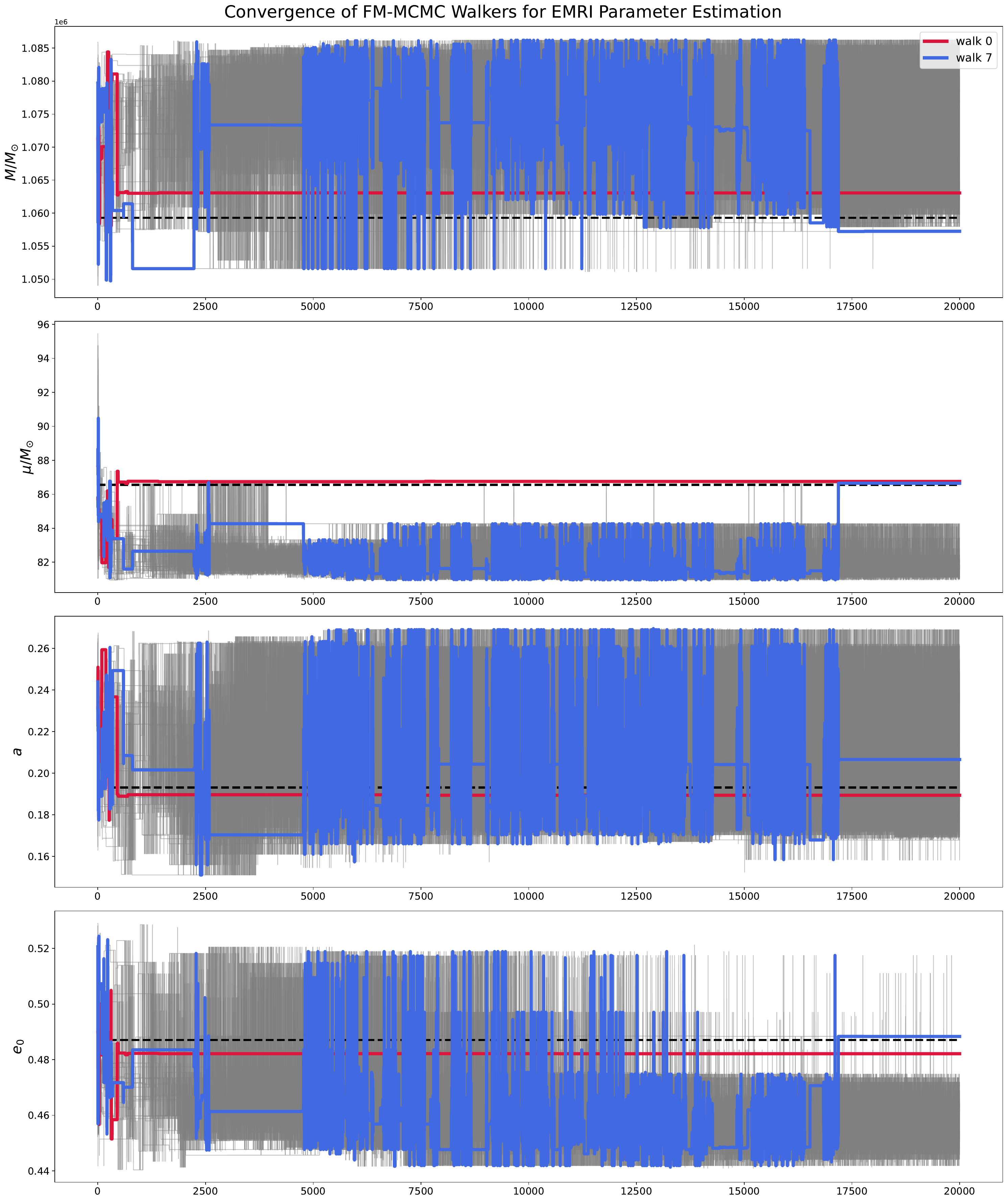}
  \caption{{This figure shows the walk results for FM-MCMC on the posterior distribution of the EMRI parameters. The plot demonstrates the multimodal nature of the posterior surface, where FM-MCMC successfully explores multiple modes of the likelihood. The walkers effectively navigate through global and local optima, highlighting FM-MCMC’s ability to explore the full parameter space.}
}
\label{A2}
\end{figure}

\begin{figure}[htb]
  \setlength{\arrayrulewidth}{0.2pt} 
  \centering
  \includegraphics[width=0.85\textwidth]{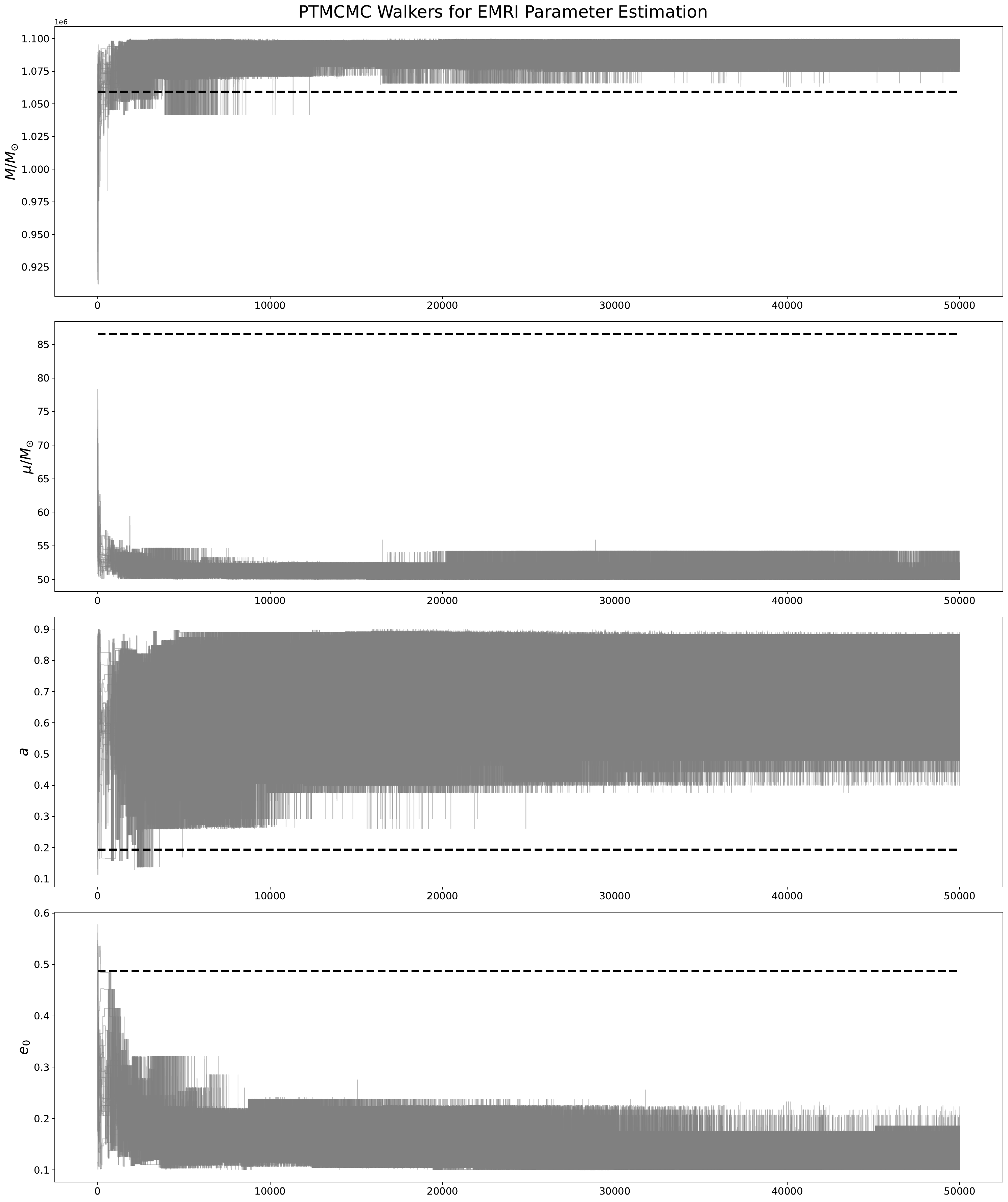}
  \caption{{This figure shows the walk results for PT-MCMC on the posterior distribution of the EMRI parameters. The plot illustrates how PT-MCMC gets trapped in local maxima and fails to explore other regions of the posterior distribution. Despite the use of parallel tempering, PT-MCMC is limited in its ability to explore multiple modes in the posterior.}
}
\label{A3}
\end{figure}

{
In this appendix, we also present additional walk results for both FM-MCMC and PT-MCMC methods to demonstrate the multimodal nature of the posterior distribution more clearly. As shown in Figure~\ref{A2} and Figure~\ref{A3}, the posterior distributions exhibit multiple modes, which are explored more effectively by FM-MCMC compared to PT-MCMC.}

\section{Comparison with Neural Posterior Estimation}
\begin{figure}[htb]
  \setlength{\arrayrulewidth}{0.2pt} 
  \centering
  \includegraphics[width=0.75\textwidth]{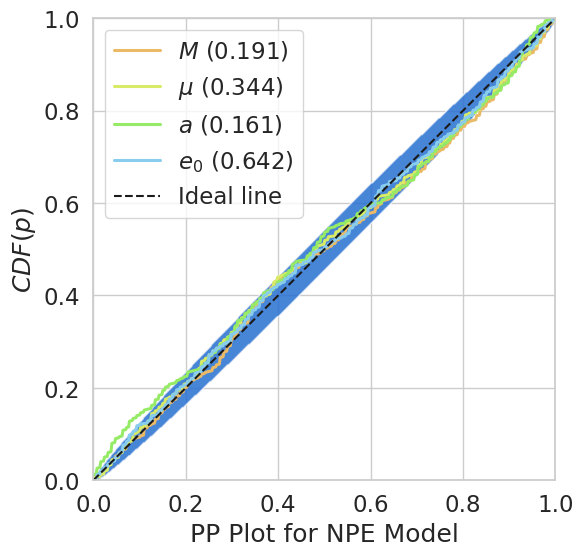}
  \caption{{
PP plot for NPE model, based on 500 simulated EMRI signals. The dashed linerepresents the ideal case, while the colored lines represent the empirical CDFs of the parameters.
The shaded blue region indicates the 95\% confidence bands. }
}
\label{A4}
\end{figure}

{
In addition to the main comparisons with Eryn, we further evaluate our FMPE framework against Neural Posterior Estimation (NPE), which represents another widely used class of neural posterior estimation techniques. To ensure fairness, both FMPE and NPE were implemented with the same base neural architecture. However, a crucial distinction is that NPE requires neural spline flows for density estimation. In our setup, we employed six neural spline flows, which significantly increases the number of trainable parameters in the model. As a result, the model size of NPE is approximately 300 million parameters, while FMPE contains only about 100 million parameters. In terms of training cost, NPE requires around three days of training on a single NVIDIA RTX 4090 GPU, whereas FMPE converges within about 20 hours under the same hardware conditions. The calibration performance is further demonstrated by the P–P plots in Figure~\ref{fig:pp} (FMPE) and Figure~\ref{A4} (NPE). While both methods yield statistically valid posterior samples, the average p-values across all intrinsic parameters are consistently higher for FMPE than for NPE, indicating better overall calibration. }

\section{Results under Fully Uninformed Initialization}
{
In this appendix, we present the results of EMRI parameter estimation when all eight parameters, including both intrinsic parameters ($M$, $\mu$, $a$, $e_0$) and extrinsic parameters ($\theta_S$, $\phi_S$, $\theta_K$, $\phi_K$), are initialized under fully uninformed priors. Specifically, the MCMC chains are initialized by random sampling across the entire prior ranges as defined in Table 1, without any constraints near the injected true values. 
This setup provides a stringent test of the robustness of our method, as it reflects the most challenging scenario for realistic searches in which no prior information is assumed. 
To ensure a fair comparison, both FM-MCMC and PT-MCMC were run with identical parallel tempering configurations: 20 walkers and 20 temperature ladder levels.}
{
We include detailed convergence diagnostics for both FM-MCMC and PT-MCMC, showing walker traces for all eight parameters. In addition, we present the posterior distributions of the intrinsic parameters under this setting. The results clearly demonstrate that FM-MCMC achieves robust convergence to the true posterior modes, even when initialized from completely uninformed states. By contrast, PT-MCMC fails to consistently escape local optima, leading to biased or unconverged results. These findings confirm the robustness of FM-MCMC under fully uninformed initialization and further validate the conclusions drawn in the main text.
}
{
We note that while FM-MCMC is capable of handling intrinsic parameters effectively in this setting, accurate recovery of extrinsic parameters may require further developments. In particular, more advanced neural network architectures, fine-tuning strategies, or hierarchical inference approaches may be necessary to improve convergence for spin polar angle  parameters. 
Exploring these directions will be an important focus of future work. 
}

\begin{figure}[htbp]
    \centering
    \includegraphics[width=1.0\textwidth]{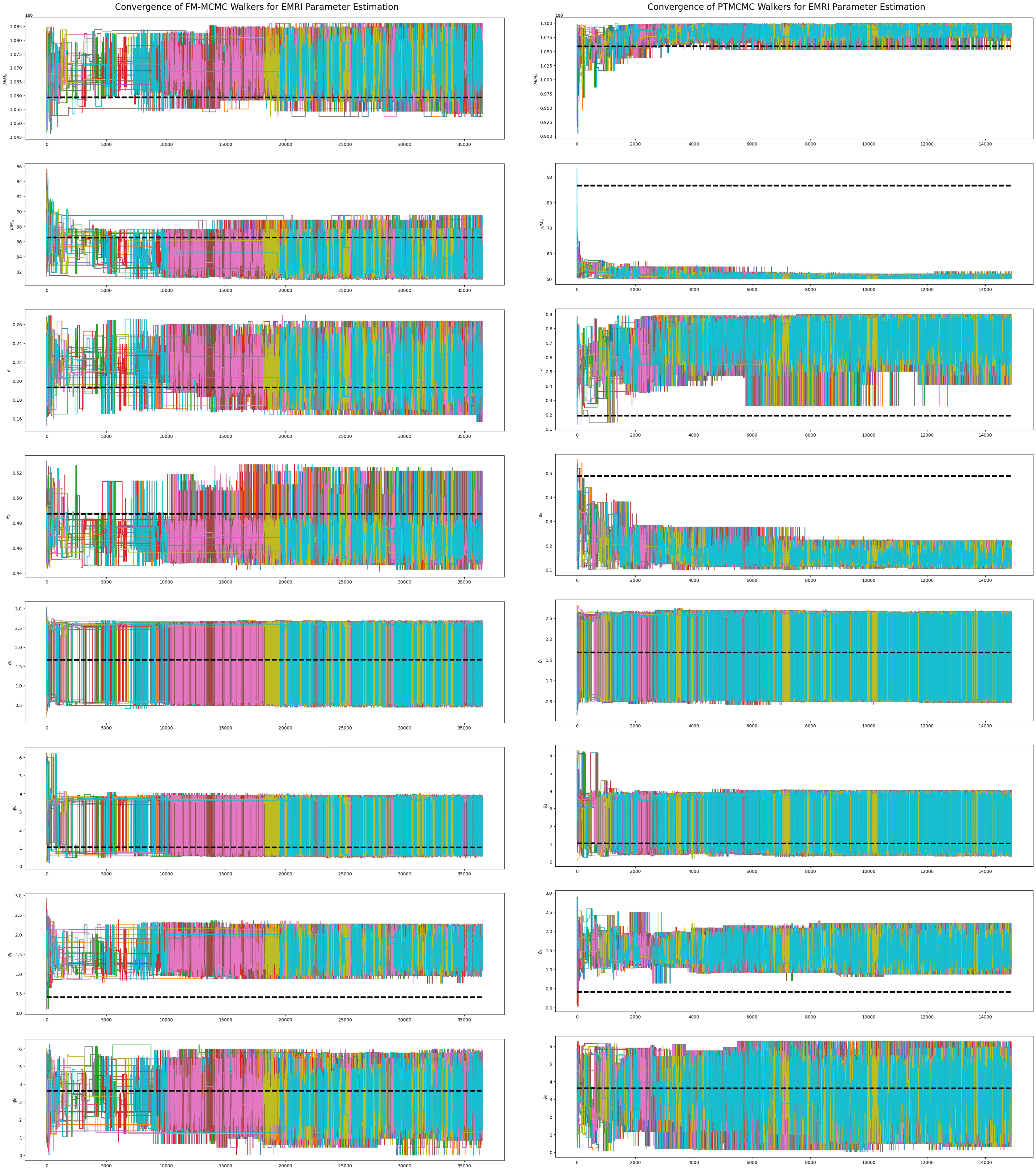}
    \caption{
    {
\textbf{ FM-MCMC vs. PT-MCMC under fully uninformed initialization.}
Convergence diagnostics and posterior distributions under fully uninformed initialization.
All eight parameters, including both intrinsic parameters ($M$, $\mu$, $a$, $e_0$) and extrinsic parameters ($\theta_S$, $\phi_S$, $\theta_K$, $\phi_K$), are initialized by random sampling across their full prior ranges as defined in Table~\ref{table:priors}.
(Left) For FM-MCMC, the walker traces show rapid convergence toward the true posterior modes, and the corresponding posterior distributions of the intrinsic parameters accurately recover the injected values.
(Right) For PT-MCMC, the chains remain trapped in local optima, and the posterior distributions exhibit significant bias, demonstrating its difficulty in exploring the multimodal landscape.
These results validate the robustness of FM-MCMC under the most challenging initialization scenario, while highlighting the limitations of PT-MCMC in fully uninformed searches.
    }}
\label{fig:appendixC_walkers_posteriors}
\end{figure}

\begin{figure}[htbp]
    \centering
    \includegraphics[width=1.0\textwidth]{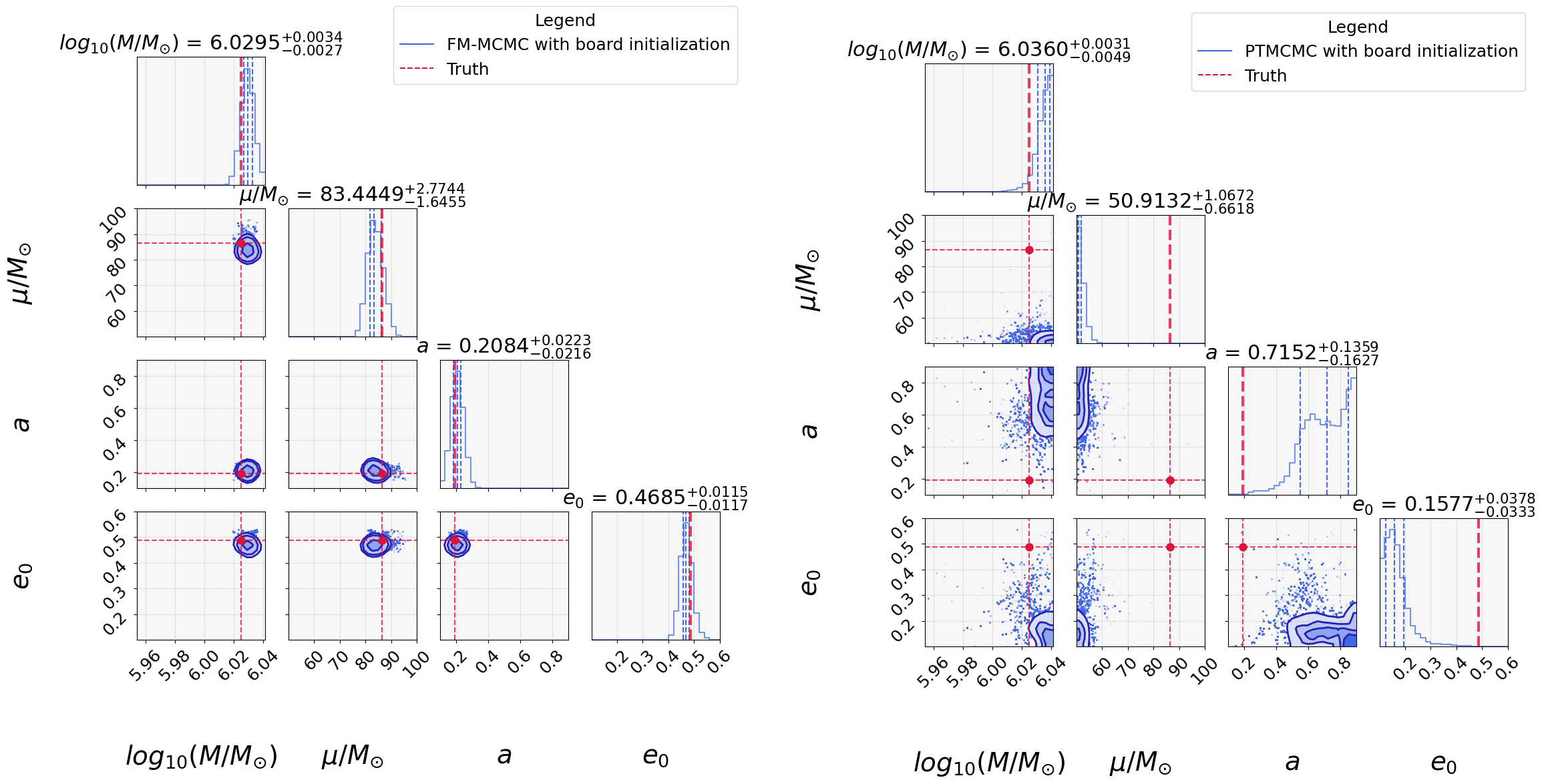}
    \caption{
    {
\textbf{ Comparison of posterior distributions under fully uninformed initialization. }
(Left) FM-MCMC successfully recovers the injected values of the intrinsic parameters ($M, \mu, a, e_0$) with tight posterior distributions, demonstrating robust convergence to the true global modes.
(Right) PTMCMC results show significant bias and broader distributions, indicating that the sampler remains trapped in local optima.
These results highlight the superior ability of FM-MCMC to explore the multimodal likelihood surface compared to PTMCMC.}
    }
\label{fig:appendixC_posteriors}
\end{figure}

\section{Network Implementation}
{
The CNF is trained on a dataset of 20,000 simulated EMRI waveforms. During training, independent Gaussian noise realizations are dynamically added to each clean signal at every epoch, ensuring that the model learns to generalize across varying noise conditions rather than overfitting to specific noise samples. Each waveform corresponds to a two-month Taiji observation sampled at 0.1 Hz, yielding a time series of approximately 200,000 data points per channel.}

{
The feature extraction network preceding the flow model is implemented as a convolutional neural network specifically designed to handle long-duration gravitational-wave time series. The network begins with a large-kernel convolutional layer (kernel size 65, stride 32) that performs strong temporal downsampling while preserving key waveform features. It is followed by a stack of six residual blocks with dilation factors of ${1, 2, 4, 8, 16, 32}$, enabling the model to capture both short- and long-timescale dependencies effectively. Each residual block consists of two 1D convolutional layers (kernel size 3, dilation $d$) with identical channel dimensions, followed by Batch Normalization and GELU activation. The residual connection directly adds the block input to the output of the second convolution, ensuring stable gradient flow and preserving information across layers. The final GELU activation provides smooth nonlinear transformation, enhancing the model’s expressivity for complex waveform structures.
After the residual stack, a sequence of convolution, pooling, and fully connected projection layers further compresses the representation into a 2048-dimensional latent embedding.}

{
The second core component is the deep conditional normalizing flow network itself, which is structured as a sequence of 21 residual blocks. This network is designed to perform a progressive, hierarchical transformation of features. It begins by processing a combined input of the compressed data representation, the latent parameters, and an embedded time variable, with an initial dimension of 3072. Through its deep structure, it systematically refines and compresses this information across multiple layers, ultimately mapping it down to the 4-dimensional vector field that defines the probability flow. The architecture employs a symmetric, funnel-like design, starting with wider layers (e.g., 8192 units) to capture broad features and gradually narrowing through successive blocks (e.g., 4096, 2048, 1024 units and so forth) to distill increasingly fine-grained information, culminating in the final output dimension.}

{
This hierarchical design facilitates a synergistic optimization between the compressed signal embedding, the flow time parameter, and the dynamic state of the source parameters. The training was conducted on a single NVIDIA RTX 4090 GPU using the Adam optimizer with an initial learning rate of 1e-4, which was decayed following a cosine annealing schedule over the course of 2000 epochs. The complete training process required approximately 12 hours to complete.}

\end{document}